\documentclass[aps,prd,amssymb,amsmath,showpacs]{revtex4}
\usepackage{graphicx}

\begin{document}

\title{Macroscopic quantum tunneling  with decoherence
}
\author{Esteban Calzetta}
\affiliation{Departmento de F\'{\i}sica, Facultad de Ciencias
Exactas y Naturales, Universidad de Buenos Aires, Ciudad
Universitaria, Pabell\'on I, 1428 Buenos Aires, Argentina}
\author{Enric Verdaguer}
\affiliation{Departament de F\'{\i}sica Fonamental and CER en
Astrof\'\i sica, F\'\i sica de Part\'\i cules i Cosmologia,
Universitat de Barcelona, Av.~Diagonal 647, 08028 Barcelona,
Spain}



\begin{abstract}
The tunneling probability for a system modelling macroscopic quantum
tunneling is computed. We consider an open quantum system with one degree of
freedom consisting of a particle trapped in a cubic potential interacting
with an environment characterized by a dissipative and a diffusion
parameter. A representation based on the energy eigenfunctions of the
isolated system, \textit{i.\;e.} the system uncoupled to the environment, is
used to write the dynamical master equation for the reduced Wigner function of the
open quantum system. This equation becomes very simple in that
representation. The use of the WKB approximation for the eigenfunctions
suggests an approximation which allows an analytic computation of the
tunneling rate, in this real time formalism,
when the system is initially trapped in the false
ground state. We find that the decoherence
produced by the environment suppresses tunneling in agreement with results
in other macroscopic quantum systems with different potentials.
We confront our analytical predictions
with an experiment where the escape rate from the zero voltage state
was measured for a current-biased Josephson junction shunted with a resistor.
\end{abstract}

\maketitle


\section{Introduction}

\label{sec1}

Macroscopic quantum tunneling is a topic of interest that pertains
to the boundaries between quantum and classical physics. This
field has undergone extensive research in recent years as
experimental advances have made possible the observation of
quantum tunneling effects in some macroscopic quantum variables
such as the flux quantum transitions in a superconducting quantum
interference device, or the decay of a zero-voltage state in a
current-biased Josephson junction \cite
{CalLeg83b,DevMarCla85,MarDevCla87,CleMarCla88,WalEtAl03}.
Macroscopic quantum
systems may be modelled by open quantum systems which are
characterized by a distinguished subsystem, described by suitable
degrees of freedom which are subject to physical experimentation,
within a larger closed quantum system. The degrees of freedom of
the remaining system are not subject to experimental observation
and act as an environment or bath for the distinguished subsystem,
which is usually referred to as the ``system'' for short. The
environment acts as a source of dissipation and noise for the
system and produces quantum decoherence. Caldeira and Leggett in
two influential papers \cite{CalLeg81,CalLeg83b} considered the
effect of dissipation on the tunneling rate and noted that
dissipation always suppresses tunneling; see also Ref.
\cite{LegEtAl87}. This work was then extended to many other open
quantum systems with different system-environment couplings, and
different potentials for the field \cite
{GraWeiHan84,GraOlsWei87,Han86,Han87,WeiEtAl87,GriEtAl89,MarGra88,GraWei84};
see Refs. \cite {HanTalBor90,Wei93} for comprehensive reviews.

Most work on macroscopic quantum tunneling is based on imaginary
time formalisms such as the Euclidean functional techniques which
have been introduced in the classical field of noise-activated
escape from a metaestable state \cite{Lan67}, or the instanton
approach introduced for quantum mechanical tunneling or for vacuum
decay in field theory \cite
{VolKobOku75,Col77,CalCol77,ColGlaMar78,ColDeL80,Col85}. These
techniques are specially suited for equilibrium or near
equilibrium situations, but cannot be generalized to truly non
equilibrium situations.
To be able to deal with out of equilibrium
situations we need a real time formalism
which describes the evolution of the quantum system by means of
true dynamical equations.

There are theoretical and practical reasons for a formalism 
of nonequilibrium macroscopic quantum tunneling. On the theoretical side
dissipation, for instance, is only truly understood in a dynamical real time
formalism. In the classical context thermal activation from metaestable
states is well understood since Kramers \cite{Kra40} in terms of the
dynamical Fokker-Planck transport equation, where the roles of dissipation
and noise and their inter-relations are known. On the other hand, an open
quantum system may be described by a dynamical equation for the reduced
density matrix, the so-called master equation, or the equivalent equation
for the reduced Wigner function which has many
similarities to the Fokker-Planck equation.
However, at present no compelling derivation of the tunneling rate is
available in this dynamical framework, that might be compared to the
instanton approach for equilibrium systems. Consequently, the effect of
dissipation, noise and decoherence on tunneling and their inter-connections
is not yet fully understood. On the practical side out of equilibrium
macroscopic quantum tunneling is becoming necessary to understand arrays of
Josephson junctions, or time-dependent traps for cold atoms which are
proposed for storing quantum information in future quantum computers \cite
{ShnSchHer97,MooEtAl99,CleGel04,Mon02}, or to understand
first order phase transitions in
cosmology \cite{Kib80,RivLomMaz02}.

In this paper we propose a formulation of macroscopic quantum tunneling
based on the reduced Wigner function. In recent years we have considered
different scenarios in which metaestable quantum systems are described by
the master equation for the reduced Wigner function. By using techniques
similar to those used for thermal activation processes on metastable states
\cite{Kra40,Lan69} we were able to compute the contribution to the quantum
decay probability due to the environment. This was used in some
semiclassical cosmological scenarios for noise induced inflation \cite
{CalVer99} due to the back reaction of the inflaton field, in the context of
stochastic semiclassical gravity \cite
{CalHu94,CamVer96,CalCamVer97,MarVer99a,MarVer99b}; see Refs. \cite
{HuVer03,HuVer04} for reviews on this subject. It was also used for bubble
nucleation in quantum field theory, where the system was described by the
homogeneous mode of the field of bubble size and the environment was played
by the inhomogeneous modes of the field \cite{CalRouVer01a,CalRouVer02}, and
on some simple open quantum systems coupled linearly to a continuum of
harmonic oscillators at zero temperature \cite{ArtEtAl03}. But in all these
problems only the contribution to tunneling due to activation was
considered. The reason is that the non harmonic terms of the potential that
induce the pure quantum tunneling of the system lead to third and higher
order momentum derivatives of the reduced Wigner function in the master
equation. One has to resort to numerical methods such as those based on
matrix continued fractions in order to compute decay rates from master
equations in this case \cite{Ris89,VogRis88,RisVog88,GarZue04}.

In this paper we are able to deal simultaneously with both contributions,
namely pure quantum tunneling and thermal activation, to the vacuum decay
process using the master equation for the reduced Wigner function. The
computational innovation that makes this possible is the introduction of a
representation of the reduced Wigner function based on the energy wave
functions of the isolated system, \textit{i.\;e.} the system not coupled to
the environment. This representation is useful in a way somewhat analogous
to the way the energy representation is useful in the Schr\"{o}dinger
equation. It is quite remarkable that in this representation the master
equation can be solved analytically under certain approximations. The key to
this result is that quantum tunneling is already encoded in the energy wave
functions, which we can compute in a WKB approximation.

In order to have a working model in a form as simple as possible, but that
captures the main physics of the problem, we put by hand the effect of the
environment with some phenomenological terms that describe noise and
dissipation in a simple form. It turns out that these terms can be deduced
from microscopic physics, when the environment is made by an Ohmic
distribution of harmonic oscillators weakly coupled in thermal equilibrium
at high temperature. Thus the model is not strictly suited to describe
tunneling from the false vacuum, or zero temperature transitions. The
appropriate equations are known in this case \cite{HuPazZha92,ArtEtAl03} and
include time dependent noise and dissipation coefficients, and anomalous
diffusion. Thus the model studied here is a toy model at low temperature. We
may expect qualitative agreement with relevant experiments at very low
temperature, but no precision comparable to the instanton approach in this
case. We illustrate this by comparing the analytical predictions of our
model with an experiment where the escape rate from the zero voltage state
was measured for a current-biased Josephson junction shunted with a resistor
\cite{CleMarCla88}. Since our main purpose here is the introduction of a
working formalism for out of equilibrium macroscopic quantum tunneling we
leave the more realistic and involved computation for future work.

Master equations play also an important role in elucidating the emergence of
classicality in open quantum systems as a result of their interaction with
an environment. In fact, as the master equation gives the quantum evolution
of initial states, defined by the reduced Wigner function at some initial
time, it has been of great help to study decoherence. In particular, it has
been used to clarify the decoherence time scales, or the way in which the
environment selects a small set of states of the system which are relatively
stable by this interaction, the so-called pointer states, whereas the
coherent superposition of the remaining states are rapidly destroyed by
decoherence \cite{Zur91,PazHabZur93,ZurPaz94,PazZur99,PazZur01}. Recently
the effect of the interaction with the environment on coherent tunneling has
been analyzed in the framework of an open quantum system that is classically
chaotic: a harmonically driven quartic double well \cite{MonPaz00,MonPaz01}.
It turns out that in this problem, which requires a numerical analysis,
tunneling is suppressed as a consequence of the interaction with the
environment. The dissipation and diffusion terms in the master equation have
been derived assuming a high-temperature limit of an Ohmic environment, as
in the model discussed in the present work. Thus, our calculation can also
be understood in the context of the study of decoherence on
quantum tunneling. In fact, we are able to identify the terms responsible
for decoherence and their effect on tunneling in our model.

This paper is organized as follows. In the next two sections \ref{sec2} and
\ref{sec3} we review the theory of tunneling in closed systems and introduce the
energy representation for Wigner functions. This extended review is
necessary both to establish our conventions and to recall specific results
which are central to the main argument. In section \ref{oqs} we introduce
the environment and compute the dynamical equation for the reduced Wigner
function of the system. In section \ref{ab} we introduce the so-called
quantum Kramers equation, which is a local approximation to the transport
equation which captures the essential physics. For the purpose of computing
tunneling from the false vacuum state, the quantum Kramers equation can be
simplified further under the so-called phase shift approximation. In section
\ref{tunnel} we apply the foregoing to the actual estimate of the tunneling
amplitude for the open system. In section \ref{experiment} we compare the
analytical predictions of our model with an experiment on a current-biased
Josephson junction. Finally, in Section \ref{conclusions} we briefly
summarize our results. In the Appendixes we provide additional technical
details.

\section{Tunneling in quantum mechanics}

\label{sec2}

In this section we review the WKB method to tunneling in quantum mechanics.
The energy eigenfunctions in the WKB approximation we obtain will play an
important role in the energy representation of the Wigner function that will
be introduced latter.

\subsection{The system}

\begin{figure}[tbp]
\includegraphics[height=8cm]{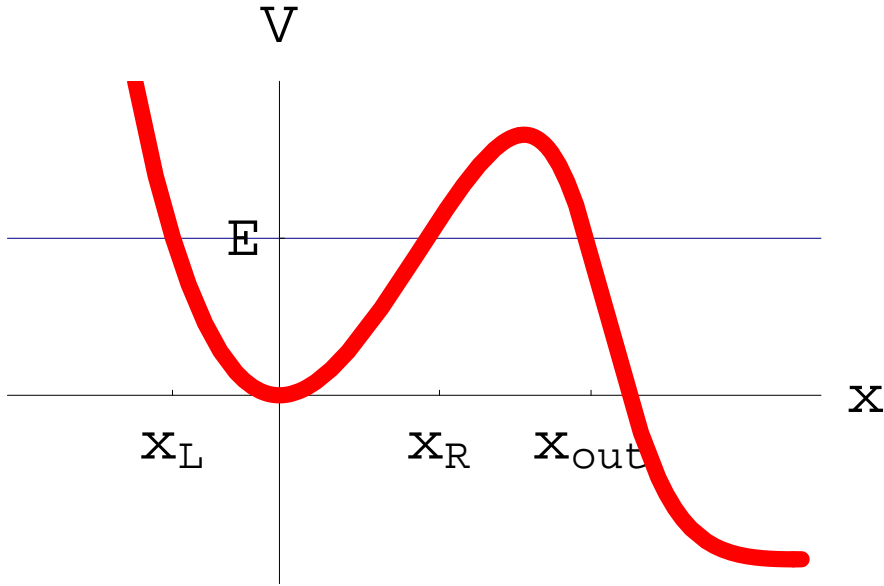}
\caption{A schematic plot of the potential. For an energy E below the
barrier there will be three classical turning points, also shown.}
\end{figure}

We begin with the simple closed quantum mechanical system formed by a
particle of mass $M$ in one dimension described by a Hamiltonian
\begin{equation}
H=\frac{p^{2}}{2M}+U\left( x\right) ,  \label{hamilton}
\end{equation}
with a potential $U$ given by
\begin{equation}
U\left( x\right) =\frac{1}{2}M\Omega _{0}^{2}x^{2}-\frac{\lambda }{6}x^{3},
\label{potential}
\end{equation}
for small values of the coordinate $x$. This is a fairly general potential
for a tunneling system, it is the basic element in the dashboard potential,
which is a very good model for a flux trapped in a superconducting quantum
interference device (SQUID), or a single Josephson junction biased by a
fixed external current \cite{CalLeg83b,MarDevCla87,Wei93,Tin96}. For
technical reasons, it is convenient to assume that for large $x$ the
potential flattens out and takes the value $U\left( x\right) =-U_{\infty },$
both negative and constant. The tunneling process ought to be independent of
the form of the potential this far away from the potential barrier. We
present a sketch of this potential in Fig. 1.

There is one classically stable point at $x=0$, and one unstable point $%
x=x_{s}=2M\Omega _{0}^{2}/\lambda ,$ \ corresponding to an energy $%
\varepsilon _{s}=2M^{3}\Omega _{0}^{6}/(3\lambda ^{2})$. The curvature of
the potential is $U^{\prime \prime }(0)=M\Omega _{0}^{2}$ at $x=0,$ and $%
-U^{\prime \prime }(0)$ \ at $x_{s}.$ The other point at which $U\left(
x\right) =0$ is $x=x_{exit}=(3/2)x_{s}.$ For $x\gg x_{exit}$ the potential
flattens out and is constant.

\subsection{The WKB approximation}

\label{wkb}

If we assume that the particle is trapped in the potential well, that is in
its false ground state or false vacuum, the tunneling probability can be
computed in this simple problem in many ways. One of the most efficient is
the instanton method which reduces to the computation of the ``bounce
solution''. The most attractive aspect of this computation is that it can be
easily extended to field theory where the tunneling probability is then
interpreted as the probability per unit time and volume to nucleate a bubble
of the true vacuum phase. The rate for quantum tunneling is $\Gamma
_{closed}=(\Omega _{0}/2\pi )a_{q}\exp (-S_{B}/\hbar )$, where $S_{B}$ is
the action for the ``bounce'' (or instanton), namely the solution to the
classical equations of motion which interpolates between $x=0$ and $%
x=x_{exit}$ in imaginary time
$S_{B}=2\int_{0}^{x_{exit}}dx\;\sqrt{2MU\left( x\right) },$ and
the prefactor $a_{q}=(120\pi S_{B}/\hbar )^{1/2}$. Our expression
for the potential is so simple that the above integral can be
computed explicitly: $S_{B}/\hbar =18\varepsilon
_{s}/(5\varepsilon _{0})$, where $\varepsilon
_{0}=\frac{1}{2}\hbar \Omega _{0}$ is the zero point energy of a
harmonic oscillator with frequency $\Omega _{0}.$

Here, however, we will concentrate on a real time approach by expanding the
false vacuum state as a linear combination of true eigenstates of the
Hamiltonian. To the required accuracy, it is enough to work with the WKB
approximations to the true eigenfunctions; see for instance Refs. \cite
{LanLif77,GalPas90}. The instanton method reviewed in the previous paragraph
can, in fact, be easily justified by this semiclassical approximation. Here
we explain in some detail this standard procedure to obtain the
eigenfunctions by matching the WKB solutions in the different regions of the
potential. These solutions will play a crucial role in the energy
representation for the Wigner functions to be introduced latter.

Let $0<E<\varepsilon _{s}$ be the energy of the particle in the potential
well, and $\psi _{E}$ the corresponding eigenfunction. The Schr\"{o}dinger
equation is
\begin{equation}
\frac{-\hbar ^{2}}{2M}\frac{\partial ^{2}}{\partial x^{2}}\psi _{E}+U\left(
x\right) \psi _{E}=E\psi _{E}.  \label{wkb1}
\end{equation}
Let us define
\begin{equation}
p\left( x\right) =\sqrt{2M\left| U\left( x\right) -E\right| },  \label{wkb2}
\end{equation}
and the integral $S(x,y)$ (note the order in the integration limits)
\begin{equation}
S\left( x,y\right) =\int_{y}^{x}dx^{\prime }\;p\left( x^{\prime }\right) .
\label{wkb3}
\end{equation}

The WKB solutions are obtained from these elements. We have to match the WKB
solutions in the different regions across the potential function. The
details of this calculation are given in Appendix A. The WKB solution $\psi
_{E}$ for energies in the range $0<E<\varepsilon _{s}$ is given by Eq. (\ref
{wkb10}), where $x_{L}<x_{R}<x_{out}$ are the three classical turning points
for the cubic potential (\ref{potential}); see Fig.~1. The normalization
constant $K_{E}$ in Eq. (\ref{wkb10}) is obtained by imposing the continuous
normalization of the eigenfunctions given in Eq. (\ref{norm1}) and it is
given in Eq. (\ref{norm6}). Of particular relevance is the value of the
eigenfunction $\psi _{E}(x)$ at values $x\gg x_{out}$. This gives the main
contribution to the continuous normalization integral. The value of the
eigenfunction at $x\gg x_{out}$, as computed in Appendix A, is
\begin{equation}
\psi _{E}(x)\sim \sqrt{\frac{2M}{\hbar \pi p_{\infty }}}\sin \left( \frac{%
p_{\infty }x}{\hbar }+\delta _{E}\right) ,  \label{norm8}
\end{equation}
where the phase $\delta _{E}$ is introduced in Eqs. (\ref{norm7}) and $%
p_{\infty }(E)$ is defined by Eq. (\ref{wkb2}) when $x\gg x_{out}$; see also
Eq. (\ref{norm2}).

We are interested in the details of the eigenfunctions near the false vacuum
state, since we will be dealing with tunneling from vacuum. Thus, in the
remaining of this section we give explicitly the values of the normalization
constant $K_{E}$ and the phase shifts $\delta _{E}$ near this vacuum state.
Therefore let us impose the Bohr-Sommerfeld quantization condition (\ref
{wkb11}) and let $E_{0}$ be the corresponding lowest energy, that is, $n=0$
in Eq. (\ref{wkb11}). As we will see in the next subsection this defines the
false vacuum energy. Expanding the integral in Eq. (\ref{wkb3}) around $E_{0}
$ we find that close to the lowest energy value
\begin{equation}
S\left( x_{R},x_{L}\right) \sim \frac{\pi \hbar }{2}-\tau \left(
E-E_{0}\right) ,  \label{norm11}
\end{equation}
where $\tau $ is defined by
\begin{equation}
\tau =\int_{x_{L}}^{x_{R}}dx\;\sqrt{\frac{2M}{U\left( x\right) -E_{0}}}.
\label{norm12}
\end{equation}
Thus $\cos ^{2}\left( S\left( x_{R},x_{L}\right) /\hbar \right) \sim (\tau
^{2}/\hbar ^{2})\left( E-E_{0}\right) ^{2}$, and evaluating the right hand
side of (\ref{norm10}) at $E_{0},$ we conclude that $K_{E}^{2}$ has poles at
the complex energies
\begin{equation}
E_{\pm }=E_{0}\pm i\varepsilon ,\qquad \varepsilon \equiv \frac{\hbar }{%
4\tau }e^{-2S_{0}\left( x_{out},x_{R}\right) /\hbar },  \label{norm13}
\end{equation}
which is in agreement with the standard result \cite{GalPas90}. To simplify
the notation let us call $S_{0}=S_{0}\left( x_{out},x_{R}\right) $ and $%
f_{0}=f\left( E_{0}\right) +\pi /4$, then we have from Eqs. (\ref{norm5a})
and (\ref{norm5b}) that the functions $A(E)$ and $B(E)$ for $E$ near $E_{0}$
are: $A\left( E\right) =(\tau /\hbar )\exp (S_{0}/\hbar )[F_{-}(E)+F_{+}(E)]$
and $B\left( E\right) =(-i\tau /\hbar )\exp (S_{0}/\hbar )[F_{-}(E)-F_{+}(E)]
$, where $F_{-}(E)=\exp (if_{0}/\hbar )\left( E-E_{-}\right) $ and $%
F_{+}(E)=\exp (-if_{0}/\hbar )\left( E-E_{+}\right) $. Notice that neither $A
$ nor $B$ vanish at $E_{\pm }$. Finally from Eq. (\ref{norm6}) we can write
the normalization constant near the false vacuum energy, as
\begin{equation}
K_{E}^{2}=\frac{M}{\pi \hbar \tau }\frac{\varepsilon }{\left( E-E_{0}\right)
^{2}+\varepsilon ^{2}}=\frac{4M\varepsilon ^{2}}{\pi \hbar ^{2}}\frac{%
e^{2S_{0}/\hbar }}{\left( E-E_{-}\right) \left( E-E_{+}\right) },
\label{norm15a}
\end{equation}
and from Eqs. (\ref{norm7}) the phase shifts are
\begin{equation}
e^{i\delta _{E}}=2\sqrt{\frac{\varepsilon ^{2}e^{2S_{0}/\hbar }}{\left(
E-E_{0}\right) ^{2}+\varepsilon ^{2}}}\left( A+iB\right) =e^{if_{0}/\hbar }%
\sqrt{\frac{E-E_{-}}{E-E_{+}}}.  \label{norm15b}
\end{equation}

Equations (\ref{norm8}), (\ref{norm15a}) and (\ref{norm15b}) are the main
results of this section. We notice, in particular, the poles of the norm and
the phase shifts at $E_\pm$ near the false vacuum energy. The strong
dependence on the energy of these functions near the false ground energy
will play an important role in the next sections.

\subsubsection{The false vacuum}

\label{vacu}

Before we start with the computation of the tunneling rate we have to define
what we mean by the decaying state, all the wave functions we considered so
far are true stationary states and, obviously, show no decay whatsoever. We
need to confine initially the particle into the potential well in its lowest
energy. To this end, we introduce an auxiliary potential $U_{\mathrm{aux}}$
which agrees with $U$ up to $x_{s}$ (where the true potential reaches its
maximum value) and increases thereafter. We may assume that the growth of $%
U_{\mathrm{aux}}$ is as fast as necessary to justify the approximations
below; the tunneling rate is insensitive to the details of $U_{\mathrm{aux}}$
beyond $x_{s}.$ Thus, we define the decaying state $\psi _{0}$ as the ground
state of a particle confined by $U_{\mathrm{aux}}$ \cite{Mig77}.

It is obvious from the form of the WKB solutions that $\psi _{0}$ agrees
with $\psi _{E_{0}}$ up to $x_{s}$, \textit{i.\ e.} $\psi _{0}(x)=\psi
_{E_{0}}(x)$ for $x\leq x_{s}$, where $E_{0}$ is the Bohr-Sommerfeld ground
state energy for the auxiliary potential $U_{\mathrm{aux}}$, which
corresponds to $n=0$ in the condition (\ref{wkb11}). Beyond $x_{s}$, $\psi
_{0}$ will decay rapidly to zero, unlike $\psi _{E_{0}}.$ Like any other
wave function, $\psi _{0}$ admits a development in the complete base of
energy eigenfunctions $\psi _{E}$, as
\begin{equation}
\psi _{0}\left( x\right) =\int dE\;C_{E}\psi _{E}\left( x\right) ,
\label{vacu1a}
\end{equation}
where due to our normalization the Fourier coefficients are given by
\begin{equation}
C_{E}=\int dx\;\psi _{E}\left( x\right) \psi _{0}\left( x\right) .
\label{vacu1b}
\end{equation}

To find these coefficients, we observe that $\psi _{0}\left( x\right) $ is a
solution to the Schr\"{o}dinger equation with the auxiliary potential $U_{%
\mathrm{aux}}$
\begin{equation}
\frac{-\hbar ^{2}}{2M}\frac{\partial ^{2}}{\partial x^{2}}\psi _{0}+U_{%
\mathrm{aux}}\left( x\right) \psi _{0}=E_{0}\psi _{0}.  \label{vacu2}
\end{equation}
Let us add to both sides of this equation the term $[U\left( x\right) -U_{%
\mathrm{aux}}\left( x\right) ]\psi _{0}$ and then multiply both sides by $%
\psi _{E}\left( x\right) $ and integrate to obtain
\begin{equation}
\left( E-E_{0}\right) C_{E}=-\int_{x_{s}}^{\infty }dx\;\psi _{E}\left(
x\right) \left[ U_{\mathrm{aux}}\left( x\right) -U\left( x\right) \right]
\psi _{0}\left( x\right) .  \label{new1}
\end{equation}
An important consideration is that $\psi _{0}\left( x\right) $ is a smooth
function (as opposed to a distribution), and, unlike $\psi _{E_{0}}$ it is
normalizable, so $C_{E}$ must also be smooth. This means that it is
allowable to assume $E\neq E_{0}$ in Eq. (\ref{new1}); $C_{E_{0}}$ can then
be found by analytical continuation. To estimate the right hand side of Eq. (%
\ref{new1}), let us introduce; cf. Eq. (\ref{wkb2}),
\begin{equation}
p_{\mathrm{aux}}\left( x\right) =\sqrt{2M\left| U_{\mathrm{aux}}\left(
x\right) -E_{0}\right| }.
\end{equation}
To the right of $x_{s}$ we may use the WKB approximation with the decaying
solution into the forbidden region to write
\begin{equation}
\psi _{0}\left( x\right) =\psi _{0}\left( x_{s}\right) e^{-\frac{1}{\hbar }%
\int_{x_{s}}^{x}p_{\mathrm{aux}}\left( y\right) dy}.  \label{vacu3}
\end{equation}

On the other hand, $\psi _{E}\left( x\right) $ is given by Eq. (\ref{eq89})
in Appendix A. If $E$ is close to $E_{0}$, then Eq. (\ref{norm11}) applies,
and we may write
\begin{equation}
\psi _{E}\left( x\right) \sim 2K_{E}\left[ \frac{\tau }{\hbar }\left(
E-E_{0}\right) F_{+}\left( x_{s},x_{R}\right) e^{\frac{1}{\hbar }%
\int_{x_{s}}^{x}p\left( y\right) dy}+\frac{1}{2}F_{-}\left(
x_{s},x_{R}\right) e^{-\frac{1}{\hbar }\int_{x_{s}}^{x}p\left( y\right)
dy}\right] .
\end{equation}
Substituting the two previous expressions into the right hand side of Eq. (%
\ref{new1}) we see that we have to compute the two following integrals,
\begin{equation}
J_{\pm }=\int_{x_{s}}^{\infty }dx\;\left[ U_{\mathrm{aux}}\left( x\right)
-U\left( x\right) \right] e^{-\frac{1}{\hbar }\int_{x_{s}}^{x}\left[ p_{%
\mathrm{aux}}\left( y\right) \pm p\left( y\right) \right] dy}.
\label{vacu3a}
\end{equation}
The integral, $J_{-}$, is dominated by the region near the lower limit,
where $p_{\mathrm{aux}}\left( x\right) $ is close to $p\left( x\right) $ and
we can write
\[
p_{\mathrm{aux}}\left( x\right) -p\left( x\right) \sim \frac{\left( p_{%
\mathrm{aux}}^{2}\left( x\right) -p^{2}\left( x\right) \right) }{2\sqrt{%
2MU\left( x_{s}\right) }}=\sqrt{\frac{M}{2U\left( x_{s}\right) }}\left[ U_{%
\mathrm{aux}}\left( x\right) -U\left( x\right) +E-E_{0}\right] ,
\]
{}from where we obtain
\begin{equation}
J_{-}=\hbar \sqrt{\frac{2U\left( x_{s}\right) }{M}}-\left( E-E_{0}\right)
\int_{x_{s}}^{\infty }dx\;e^{-\frac{1}{\hbar }\int_{x_{s}}^{x}\left[ p_{%
\mathrm{aux}}\left( y\right) -p\left( y\right) \right] dy},  \label{vacu3c}
\end{equation}
where the remaining integral is made negligible by an appropriate choice of $%
U_{\mathrm{aux}}$. For the other integral, $J_{+}$, we see that the
corresponding exponential factor in Eq. (\ref{vacu3a}) decays faster than
the exponential factor of $J_{-}$, so that the region which effectively
contributes to the integral is narrower. Since the pre-exponential factor
vanishes at the lower limit, we find $J_{+}\sim 0$. Finally, putting all
these pieces together into the right hand side of Eq. (\ref{new1}) we get to
leading order,
\[
\left( E-E_{0}\right) \left[ C_{E}+2K_{E}\psi _{0}\left( x_{s}\right) \tau
\sqrt{\frac{2U\left( x_{s}\right) }{M}}F_{+}\left( x_{s},x_{R}\right)
\right] =0,
\]
whose solution, assumed smooth, is
\begin{equation}
C_{E}=-2K_{E}\psi _{0}\left( x_{s}\right) \tau \sqrt{\frac{2U\left(
x_{s}\right) }{M}}F_{+}\left( x_{s},x_{R}\right) .  \label{vacu4}
\end{equation}
We note that $C_{E}$ is independent of the choice of $U_{\mathrm{aux}}$
beyond $x_{s}$, as it should.

Thus, we have found the false vacuum wave function in terms of the energy
wave functions of the original problem. The false ground state is a
superposition of energy eigenstates which are fine tuned in such a way as to
produce destructive interference outside the potential well. Notice that $%
C_{E}$, because of the factor $K_{E}$ \ in Eq. (\ref{vacu4}), peaks near the
energy of the false ground state, and has a strong dependence on the energy
near this ground state energy.

\subsubsection{Tunneling from the false vacuum}

Let us now compute the tunneling rate assuming that the particle is
described initially by the false ground state $\psi _{0}$. At time $t$, we
have
\begin{equation}
\psi \left( x,t\right) =\int dE\;e^{-iEt/\hbar }C_{E}\psi _{E}\left(
x\right) ,  \label{tunn1}
\end{equation}
The persistence amplitude is
\begin{equation}
\rho \left( t\right) =\int dx\;\psi _{0}^\ast \left( x\right) \psi \left(
x,t\right) =\int dE\;e^{-iEt/\hbar }C_{E}^{2}.  \label{tunn2}
\end{equation}
To perform the integration we can close the contour of integration in the
complex $E$ plane adding an arc at infinity, whereby we pick up the pole $%
E_- $ in $K_{E}^{2}$; cf. Eq. (\ref{vacu4}). Therefore $\rho(t) $ goes like
\begin{equation}
\rho(t)\sim \exp \left[ \frac{-t}{4\tau} \exp\left(-\frac{2}{\hbar}S_{0}
\left( x_{out},x_{R}\right) \right)\right],  \label{tunn3}
\end{equation}
provided $t$ is not too large. The tunneling rate for this closed system, $%
\Gamma_{closed}$, may be defined from the persistence probability $%
\rho^2(t)\sim \exp(-\Gamma_{closed}t)$, so that $\Gamma_{closed}= (1/2\tau
)\exp (-2S_{0}/\hbar )$, which agrees with the result of the bounce
solution. Note that if we take the classical lowest energy $E=0$, then $%
x_{R}=x_{L}=0$, $x_{out}=x_{exit}$, and $S_{B}=2S(x_{exit},0)$, but $S_0$
here is the action corresponding to a particle with false vacuum energy $E_0$%
, which differs from zero, consequently it differs from $S_B/2$. This
difference is accounted for by the prefactor $a_q$ in the instanton result.

An equivalent way of deriving this result is to estimate the integral by a
stationary phase approximation. We can write the integral of Eq. (\ref{tunn2}%
) as
\[
\rho(t)=\int dE\;e^{F\left[ E\right] }G\left[ E\right]
\]
where $F\left[ E\right] =-iEt/\hbar +\ln (K_{E}^{2})$ with $K_{E}^{2}$ given
by (\ref{norm15a}). We consider everything else going into $G\left[ E\right]
$ as relatively slowly varying. The stationary phase points are the roots of
$F^{\prime }=0.$ When $t\rightarrow \infty $, these roots must approach $%
E_{\pm }.$ Write e. g. $E=E_{-}+\varepsilon r$. If $\left| r\right| <1$,
then $r=(i\hbar /\varepsilon t)[1-(\hbar /2\varepsilon t)+...]$; this
approximation is consistent if $\varepsilon t/\hbar >1.$ At the stationary
point
\[
F\left[ E\right] =-i\frac{E_{-}t}\hbar +\ln \left[ t\right] + \mathrm{%
constant}.
\]
The first term accounts for the exponential decay, as $E_{-}=E_0
-i\varepsilon$ where $\varepsilon$ is given by Eq. (\ref{norm13}), in
agreement with the previous result (\ref{tunn3}). On the other hand, $%
F^{\prime \prime }\sim t^2$, so the Gaussian integral over energies
contributes a prefactor of order $t^{-1}.$

\section{Wigner function and energy representation}

\label{sec3}

A very useful description of a quantum system is that given by the Wigner
function in phase space, which is defined by an integral transform of the
density matrix \cite{Wig32,HilEtAl84}. The Wigner function for a system
described by a wave function $\psi(x)$ is
\begin{equation}
W\left( x,p\right) =\int \frac{dy}{2\pi \hbar }\;e^{ipy/\hbar }\;\psi \left(
x-\frac{y}{2}\right) \psi ^{\ast }\left( x+\frac{y}{2}\right),
\label{wigner1}
\end{equation}
where the sign convention is chosen so that a momentum eigenstate $\psi
_{p_{0}}\left( x\right) \sim e^{ip_{0}x/\hbar }/\sqrt{2\pi \hbar } $ becomes
$W_{p_{0}}\left( x,p\right) =(1/2\pi \hbar) \delta \left( p-p_{0}\right) $.
Moreover, it satisfies
\begin{equation}
\int dp\;W\left( x,p\right) =\;\left| \psi \left( x\right) \right|
^{2},\qquad \int dx\;W\left( x,p\right) =\;\left| \int dx\;\frac{%
e^{-ipx/\hbar }}{\sqrt{2\pi \hbar }}\psi \left( x\right) \right| ^{2},
\label{wigner2}
\end{equation}
and it is normalized so that $\int \int dx\;dp\;W(x,p)=1$. Thus the Wigner
function is similar in some ways to a distribution function in phase space,
it is real but, unlike a true distribution function, it is not positive
defined; this is a feature connected to the quantum nature of the system it
describes.

The Schr\"{o}dinger equation for the wave function $\psi $,
\begin{equation}
\frac{-\hbar ^{2}}{2M}\frac{\partial ^{2}}{\partial x^{2}}\psi +U\left(
x\right) \psi =i\hbar \frac{\partial }{\partial t}\psi ,  \label{wigner3}
\end{equation}
translates into a dynamical equation for the Wigner function, which is
easily derived. In fact, by taking the time derivative of (\ref{wigner1}),
using the Schr\"{o}dinger equation (\ref{wigner3}), and integrating by parts
we have
\begin{eqnarray}
\frac{\partial }{\partial t}W\left( x,p\right)  &=&-\frac{i}{\hbar }\int
\frac{dy}{2\pi \hbar }\;e^{ipy/\hbar }\;\left\{ \left( \frac{-i\hbar p}{M}%
\right) \frac{\partial }{\partial x}\left[ \psi \left( x-\frac{y}{2}\right)
\psi ^{*}\left( x+\frac{y}{2}\right) \right] \right.   \nonumber \\
&&\left. +\psi \left( x-\frac{y}{2}\right) \left[ U\left( x-\frac{y}{2}%
\right) -U\left( x+\frac{y}{2}\right) \right] \psi ^{*}\left( x+\frac{y}{2}%
\right) \right\} .  \nonumber
\end{eqnarray}
For the cubic potential (\ref{potential}) we have $U\left( x-y/2\right)
-U\left( x+y/2\right) =-M\Omega _{0}^{2}xy+(\lambda /2)x^{2}y+(\lambda
/24)y^{3}$ and, noting that $ye^{ipy/\hbar }=-i\hbar \partial
_{p}e^{ipy/\hbar }$ and $y^{3}e^{ipy/\hbar }=i\hbar ^{3}\partial
_{p}^{3}e^{ipy/\hbar }$, we get the equation for the Wigner function
\begin{equation}
\frac{\partial }{\partial t}W\left( x,p\right) =\left[ U^{\prime }\left(
x\right) \frac{\partial }{\partial p}-\frac{p}{M}\frac{\partial }{\partial x}%
+\frac{\lambda }{24}\hbar ^{2}\frac{\partial ^{3}}{\partial p^{3}}\right]
W\left( x,p\right) ,  \label{wigner4}
\end{equation}
which may be interpreted as a quantum transport equation. The first two
terms on the right hand side are just the classical Liouville terms for a
distribution function, the term with the three momentum derivatives is
responsible for the quantum tunneling behavior of the Wigner function in our
problem. A theorem by Pawula \cite{Ris89} states that a transport equation
should have up to second order derivatives at most, or else an infinite
Kramers-Moyal expansion, for non-negative solutions $W(x,p,t)$ to exist. The
above equation for the Wigner function circumvents the implications of the
theorem since it need not be everywhere-positive. Even if we have an
everywhere-positive Gaussian Wigner function at the initial time, the
evolution generated by an equation such as Eq. (\ref{wigner4}) will not keep
it everywhere-positive. Thus, here we see the essential role played by the
non-positivity of the Wigner function in a genuinely quantum aspect such as
tunneling.

\subsection{The energy representation}

Given that a wave function $\psi $ can be represented in terms of the energy
eigenfunctions $\psi_E$, defined by Eq. (\ref{wkb1}), as
\begin{equation}
\psi \left( x\right) =\int dE\;C_{E}\psi _{E}\left( x\right),  \label{ener1}
\end{equation}
we can introduce a corresponding representation for $W(x,p)$ in terms of a
base of functions $W_{E_{1}E_{2}}(x,p)$ in phase space defined by
\begin{equation}
W_{E_{1}E_{2}}\left( x,p\right) =\int \frac{dy}{2\pi \hbar }\;e^{ipy/\hbar
}\;\psi _{E_{1}}\left( x-\frac{y}{2}\right) \psi_{E_{2}}^\ast \left( x+\frac{%
y}{2} \right).  \label{ener3}
\end{equation}
Then $W\left( x,p\right)$ can be written as
\begin{equation}
W\left( x,p\right) =\int dE_{1}dE_{2}\;C_{E_{1}E_{2}}W_{E_{1}E_{2}}\left(
x,p\right),  \label{ener2}
\end{equation}
where, in this case, we have $C_{E_{1}E_{2}}=C_{E_{1}}C_{E_{2}}^{\ast }$. On
the other hand from the definition of $W_{E_{1}E_{2}}(x,p)$ we can write
\[
\int \frac{dxdp}{\hbar }W_{E_{1}E_{2}}^{*}\left( x,p\right) W_{E_{1}^{\prime
}E_{2}^{\prime }}\left( x,p\right) =\int \frac{dxdy}{2\pi \hbar ^{2}}\left\{
\psi _{E_{1}}\left( x-\frac{y}{2}\right) \psi _{E_{2}}\left( x+\frac{y}{2}%
\right) \psi _{E_{1}^{\prime }}\left( x-\frac{y}{2}\right) \psi
_{E_{2}^{\prime }}\left( x+\frac{y}{2}\right) \right\},
\]
where the $p$ integration has been performed. If we now call $z=x-y/2$, $%
z^{\prime }=x+y/2$; then $dxdy=dzdz^{\prime }$, and
\begin{equation}
\int \frac{dxdp}{\hbar }W_{E_{1}E_{2}}^{\ast }\left( x,p\right)
W_{E_{1}^{\prime }E_{2}^{\prime }}\left( x,p\right) =\frac{1}{2\pi \hbar ^{2}%
}\delta \left( E_{1}-E_{1}^{\prime }\right) \delta \left(
E_{2}-E_{2}^{\prime }\right),  \label{ener4}
\end{equation}
which gives the orthogonality properties of the functions $W_{E_1 E_2}$.
This suggests that any Wigner function may be written in this basis as
\begin{equation}
W\left( x,p,t\right) =\int dE_{1}dE_{2}\;C_{E_{1}E_{2}}\left( t\right)
W_{E_{1}E_{2}}\left( x,p\right).  \label{ener5}
\end{equation}

We call this the energy representation of the Wigner function. In this
representation, the master equation or the quantum transport equation (\ref
{wigner4}) is very simple
\begin{equation}
\frac{\partial }{\partial t}C_{E_{1}E_{2}}\left( t\right) =\frac{ -i}{\hbar }
\left( E_{1}-E_{2}\right) C_{E_{1}E_{2}}\left( t\right),  \label{ener6}
\end{equation}
as one can easily verify. One might give an alternative derivation of the
tunneling rate from this equation, by taking the initial condition for the
Wigner function which corresponds to the false vacuum. In fact, in the next
section we will use the energy representation of the Wigner function to
compute this rate in a more complex problem involving coupling to an
environment. Note that the dynamics of the transport equation in the energy
representation is trivial and the initial condition is given in terms of the
coefficients (\ref{vacu4}) which we have already computed. The task would be
more difficult starting from the transport equation in phase space, such as
Eq. (\ref{wigner4}), since the third derivative term makes the solution of
the equation very complicated. One has to resort to methods such as those
based on matrix continued fractions in order to compute decay rates from
master equations for open quantum systems with third order derivative terms
\cite{Ris89,VogRis88,RisVog88,GarZue04}. We call the attention to the
similarity of this representation to that based in Floquet states that one
can use when the Hamiltonian is periodic \cite
{Shi65,MilWya83,BluEtAl91,UteDitHan94}. The power of this representation
will be seen in the following sections when we consider our quantum system
coupled to an environment.

\section{The open quantum system}

\label{oqs}

So far we have considered a simple closed quantum system. From now on we
will consider an open quantum system by assuming
that our system of interest is coupled
to an environment. As emphasized by Caldeira and Leggett \cite{CalLeg83b}
any quantum macroscopic system can be modelled by an open quantum system by
adjusting the coupling of the system and environment variables and by
choosing appropriate potentials. One of the main effects of the environment
is to induce decoherence into the system which is a basic ingredient into
the quantum to classical transition \cite
{CalLeg83b,Zur91,PazHabZur93,ZurPaz94,PazZur99,PazZur01}.

The standard way in which the environment is introduced
is to assume that the system
is weakly coupled to a continuum set of harmonic oscillators, with a certain
frequency distribution. These oscillators represent degrees of freedom to
which some suitable variable of the quantum system is coupled. One usually
further assumes that the environment is in thermal equilibrium and that the
whole system-environment is described by
the direct product of the density matrices of the system and
the environment at the initial time.
The macroscopic quantum system is then described 
by the reduced density matrix, or equivalently,
by the reduced Wigner function of the open quantum system. This latter
function is defined from the system-environment Wigner function after
integration of the environment variables.

In order to have a working model in a form as simple as possible, but
that captures the main effect of the environment, we will assume
that the reduced Wigner function, which we still call $W(x,p)$,
satisfies the following dynamical
equation,
\begin{equation}
\frac{\partial }{\partial t}W\left( x,p\right) =\left[ U^{\prime }\left(
x\right) \frac{\partial }{\partial p}-\frac{p}{M}\frac{\partial }{\partial x}%
+\frac{\lambda }{24}\hbar ^{2}\frac{\partial ^{3}}{\partial p^{3}}+\gamma
\frac{\partial }{\partial p}\left( p+M\sigma ^{2}\frac{\partial }{\partial p}%
\right) \right] W\left( x,p\right) ,  \label{open1}
\end{equation}
where $\gamma $ which has units of inverse time is the dissipation
parameter, and $\sigma ^{2}$ the diffusion coefficient.
The last two terms of this equation represent the effect of the
environment: the first describes the dissipation produced into the system
and the second is the diffusion or noise term. An interesting limit,
the so-called weak dissipation limit, is obtained when $\gamma \rightarrow 0$, so
that there is no dissipation, but the diffusion coefficient $\gamma \sigma
^{2}$ is kept fixed.
We will
generally refer to equation (\ref{open1}) as the quantum Kramers equation, or
alternatively, as the quantum transport equation.
It is worth to notice that this equation
reduces to a classical Fokker-Planck transport equation when $\hbar =0$: it
becomes Kramer's equation \cite{Kra40,Lan69} for a statistical system
coupled to a thermal bath and has the right stationary solutions. 

This equation can be derived in the high temperature limit
\cite {CalLeg83a,UnrZur89,HuPazZha92,HuPazZha93,HalYu96,CalRouVer03}.
In fact, assuming the
so-called Ohmic distribution for the frequencies of the harmonic oscillators
one obtains that, in this limit, 
$\sigma ^{2}=k_{B}T$, where $k_{B}$ is Boltzmann's
constant and $T$ the bath temperature.
In the low temperature limit, however, the master equation for
the reduced Wigner function is more
involved \cite{HuPazZha92,ArtEtAl03}. It
contains time dependent coefficients and an anomalous
diffusion term of the type $\nu\partial_x\partial_p\, W(x,p)$, where $\nu$ is the
anomalous diffusion coefficient. Nevertheless, a good approximation to
the $\sigma$ coefficient is given at zero
temperature by
$\sigma ^{2}\sim \frac{1}{2}\hbar \Omega _{0}$. For simplicity
we  will base our analysis
in that equation even though we are interested in quantum tunneling
from vacuum  which means that our quantum system is at zero temperature.

Equation (\ref{open1}) is often used to
describe the effect of decoherence produced by the diffusion coefficient to
study the emergence of classical behavior in quantum systems; this is a
topic of recent interest; see Ref. \cite{PazZur01} for a review. Of
particular relevance to our problem is the study of decoherence in quenched
phase transitions \cite{AntLomMon01}, and the effect of decoherence in
quantum tunneling in quantum chaotic systems \cite{MonPaz00,MonPaz01}.

The reduced Wigner function $W(x,p)$ describes the quantum state of the open
quantum system, and given a dynamical variable $A(x,p)$ associated to the
system its expectation value in that quantum state is defined by,
\begin{equation}
\left\langle A\left( x,p\right) \right\rangle =\int dxdp\;A\left( x,p\right)
\;W\left( x,p\right) .  \label{open2}
\end{equation}
Then one can easily prove from Eq. (\ref{open1}) that defining,
\begin{equation}
N=\int dxdp\;W\left( x,p\right) ,\qquad \left\langle E\right\rangle =\int
dxdp\;\left( \frac{p^{2}}{2M}+U\left( x\right) \right) W\left( x,p\right) ,
\label{open3a}
\end{equation}
we have $\dot{N}=0$ and $\langle \dot{E}\rangle =-\gamma (\langle
p^{2}/M\rangle -N\sigma ^{2})$.

\subsection{Energy representation of the reduced Wigner function}

Let us now use the base of functions in phase space $W_{E_{1}E_{2}}(x,p)$,
introduced in Eq. (\ref{ener3}), to represent the reduced Wigner function $%
W\left( x,p,t\right) $ as in Eq. (\ref{ener5}). The previous $N$ and $%
\left\langle E\right\rangle $ have very simple expressions in the energy
representation:
\begin{equation}
N=\int dE\;C_{EE}\left( t\right) ,\qquad \left\langle E\right\rangle =\int
dE\;E\;C_{EE}\left( t\right) .  \label{open4}
\end{equation}
To check the last equation we note that $\int dxdp\;\left[
(p^{2}/2M)+U\left( x\right) \right] W_{E_{1}E_{2}}\left( x,p\right)
=E_{1}\delta \left( E_{1}-E_{2}\right) ,$ which can be easily proved by
explicit substitution of the definition of $W_{E_{1}E_{2}}$, and trading
powers of $p$ by derivatives with respect to $y$ into expressions (\ref
{ener3}), and partial integrations.

The quantum transport equation (\ref{open1}) in the energy representation
becomes,
\begin{equation}
\frac{\partial }{\partial t}C_{E_{1}E_{2}}\left( t\right) =\frac{ -i }{\hbar
}\left( E_{1}-E_{2}\right) C_{E_{1}E_{2}}\left( t\right) +\gamma \int
dE_{1}^{\prime }dE_{2}^{\prime }\;Q_{E_{1}E_{2},E_{1}^{\prime }E_{2}^{\prime
}}C_{E_{1}^{\prime }E_{2}^{\prime }}\left( t\right),  \label{open5}
\end{equation}
where, after one integration by parts,
\begin{equation}
Q_{E_{1}E_{2},E_{1}^{\prime }E_{2}^{\prime }}= -2\pi \hbar ^{2}\int \frac{%
dxdp}{\hbar }\left( \frac{\partial }{\partial p}W_{E_{1}E_{2}}^{*}\left(
x,p\right) \right) \left( \frac{p}{M}+k_{B}T\frac{\partial }{\partial p}%
\right) W_{E_{1}^{\prime }E_{2}^{\prime }}\left( x,p\right),  \label{open6}
\end{equation}
which has the contributions from the dissipative and the diffusion or noise
parts, respectively, as
\begin{equation}
Q_{E_1E_2,E_1^{\prime }E_2^{\prime }}= Q_{E_1E_2,E_1^{\prime }E_2^{\prime
}}^{\left( D\right) }+ Q_{E_1E_2,E_1^{\prime }E_2^{\prime }}^{\left(
N\right) }.  \label{open6a}
\end{equation}
{}From Eq. (\ref{ener3}) it is easy to see that these coefficients can all
be written in terms of the following matrix elements:
\begin{eqnarray}
X_{E_1E_2}&=&\int dx\;x\;\psi _{E_1}\left( x\right) \psi _{E_2}\left(
x\right),  \label{open7a} \\
P_{E_1E_2}&=& \frac \hbar i\int dx\;\;\psi _{E_1}\left( x\right) \frac
\partial {\partial x}\psi _{E_2}\left( x\right) ,  \label{open7b} \\
\left( XP\right) _{E_1E_2}&=& \frac \hbar i\int dx\;\;x\psi _{E_1}\left(
x\right) \frac \partial {\partial x}\psi _{E_2}\left( x\right),
\label{open7c} \\
X_{E_1E_2}^2&=&\int dx\;x^2\;\psi _{E_1}\left( x\right) \psi _{E_2}\left(
x\right).  \label{open7d}
\end{eqnarray}
Explicitly, we have that
\begin{equation}
Q_{E_1E_2,E_1^{\prime }E_2^{\prime }}^{\left( D\right) }=\frac{-i}{2M\hbar }%
\left[ \left( XP\right) _{E_1E_1^{\prime }}\delta \left( E_2-E_2^{\prime
}\right) -P_{E_1E_1^{\prime }}X_{E_2E_2^{\prime }}-X_{E_1E_1^{\prime
}}P_{E_2E_2^{\prime }}+\left( XP\right) _{E_2E_2^{\prime }}\delta \left(
E_1-E_1^{\prime }\right) \right],  \label{open8a}
\end{equation}
\begin{equation}
Q_{E_1E_2,E_1^{\prime }E_2^{\prime }}^{\left( N\right) }=\frac{k_BT}{\hbar ^2%
}\left[ 2X_{E_1E_1^{\prime }}X_{E_2E_2^{\prime }}-X_{E_1E_1^{\prime
}}^2\delta \left( E_2-E_2^{\prime }\right) -X_{E_2E_2^{\prime }}^2\delta
\left( E_1-E_1^{\prime }\right) \right].  \label{open8b}
\end{equation}

Thus, in terms of the coefficients $C_{E_{1}E_{2}}$ the dynamics of the
quantum transport equation is very simple. This equation, in fact, resembles
a similar equation when a Floquet basis of states are used \cite
{Shi65,MilWya83,BluEtAl91,UteDitHan94}, which are very useful when the
Hamiltonian of the system is periodic in time. The Floquet basis is discrete
in such a case and a numerical evaluation of the corresponding matrix
elements (\ref{open7a})-(\ref{open7d}) can be performed; see for instance
\cite{MonPaz00,MonPaz01} for a recent application. It is remarkable that in
our case approximated analytic expressions for these matrix elements can be
found.

\subsection{Some properties of the matrix elements}

The matrix elements (\ref{open7a})-(\ref{open7d}) have a clear physical
interpretation and several relations can be derived among them. Note that $%
X_{E_1E_2}$ is the matrix element of the position operator $X$ in the energy
representation. Since $X\psi _E\left( x\right) =x\psi _E\left( x\right), $
we must have $\int dE_1\;X_{EE_1}\psi _{E_1}\left( x\right) =x\psi _E\left(
x\right). $

On the other hand, $P_{E_1E_2}$ is the matrix element for the momentum
operator. The canonical commutation relation $\left[ P,X\right] =-i\hbar $,
implies $\left[ H,X\right] =(-i\hbar/M)P$, and taking matrix elements on
both sides we have
\begin{equation}
\left( E_1-E_2\right) X_{E_1E_2}=-\frac{i\hbar }MP_{E_1E_2}.  \label{prop2}
\end{equation}

Also, $X_{E_1E_2}^2$ is the matrix element of $X^2$, therefore
\begin{equation}
X_{E_1E_2}^2=\int dE\;X_{E_1E}X_{EE_2}.  \label{prop3}
\end{equation}
On the other hand, $\left( XP\right) _{E_1E_2}$ is the matrix element of $XP$%
, consequently $\left[ \left( XP\right) _{E_2E_1}\right] ^{*}=-\left(
XP\right) _{E_2E_1}$ corresponds to $PX$, and $\left( XP\right)
_{E_1E_2}+\left( XP\right) _{E_2E_1}=\left[ X,P\right] _{E_1E_2}=i\hbar
\delta \left( E_1-E_2\right) $. Also $\left( XP\right) _{E_1E_2}-\left(
XP\right) _{E_2E_1} =(iM/\hbar) \left( E_1-E_2\right) X_{E_1E_2}^2 $, where
the commutator $[H,X^2 ]$ has been used in the last step, therefore
\begin{equation}
\left( XP\right) _{E_1E_2}=\frac{iM}{2\hbar }\left( E_1-E_2\right)
X_{E_1E_2}^2+\frac{i\hbar }2\delta \left( E_1-E_2\right).  \label{prop4}
\end{equation}
We have, also, that $\left( XP\right) _{E_1E_2}=\int dE\;X_{E_1E}P_{EE_2} $.
One may check, for consistency, that these relations imply $\dot N=0.$ In
Appendix B a test of the quantum transport equation in the energy
representation (and of the above matrix element properties) is given by
checking that a stationary solution with a thermal spectrum is, indeed, a
solution in the high temperature limit.

\subsection{Computing the matrix elements}

\label{computmatrix}

The matrix elements contain singular parts coming from the integrals over
the unbound region beyond $x_{s}.$ These singular parts are easy to compute,
since far enough the wave functions assume the simple form (\ref{norm8}).
When performing the calculation of the singular parts of the matrix elements
we will use that when $\bar x\rightarrow \infty$, we have the identities
\begin{equation}
\frac{\sin (p\bar x/\hbar)}{\pi p}\rightarrow \delta \left( p\right) ,
\qquad \frac{\cos (p\bar x/\hbar)}p\rightarrow 0,  \label{comput1}
\end{equation}
which can be easily checked by taking the Fourier transforms of these
functions with respect to $p$.

The computation of the singular parts of the matrix elements (\ref{open7a})-(%
\ref{open7d}) may be reduced to the evaluation of three basic integrals.
These integrals are
\begin{equation}
A_{A,S}\left( p_{1},p_{2}\right) =\int dx\;\sin \left[ \left( p_{1}\mp
p_{2}\right) x/\hbar+\delta _{1}\mp \delta _{2}\right],  \label{comput2a}
\end{equation}
and
\begin{equation}
B\left( p_{1},p_{2}\right) =\int dx\;\sin \left( p_{1}x/\hbar+\delta
_{1}\right) \sin \left( p_{2}x/\hbar+\delta _{2}\right),  \label{comput2b}
\end{equation}
where, for simplicity, we have written $p_{i}\equiv p_{\infty }\left(
E_{i}\right) $ and $\delta _{i}\equiv\delta \left( E_{i}\right) $ ($i=1,2$).
The matrix element $X_{E_{1}E_{2}}$ is
\begin{eqnarray}
X_{E_{1}E_{2}} &\sim& \frac{2M}{\hbar \pi \sqrt{p_{1}p_{2}}}\int dx\;x\;\sin
\left( p_{1}x/\hbar+\delta _{1}\right) \sin \left( p_{2}x/\hbar+\delta
_{2}\right)  \nonumber \\
\ &=&\frac{M}{\pi \sqrt{p_{1}p_{2}}}\left[ -\frac{\partial A}{\partial p_{1}}%
-\frac{\partial \tilde{A}}{\partial p_{2}}-\left( \frac{\partial \delta _{1}%
}{\partial p_{1}}+\frac{\partial \delta _{2}}{\partial p_{2}}\right)
B\right],  \label{comput3a}
\end{eqnarray}
where $A\equiv (A_S-A_A)/2$ and $\tilde{A}\equiv (A_S+A_A)/2$. The matrix
element $X_{E_{1}E_{2}}^2$ is
\begin{eqnarray}
X_{E_1E_2}^2 &\sim &\frac{2M}{\hbar \pi \sqrt{p_1p_2}}\int dx\;x^2\;\sin
\left( p_1x/\hbar+\delta _1\right) \sin \left( p_2x/\hbar+\delta _2\right)
\nonumber \\
&=&\frac{2M}{\pi \sqrt{p_1p_2}}\left[ -\frac{\partial C}{\partial p_1}%
-\left( \frac{\partial \delta _1}{\partial p_1}\right) D\right],
\label{comput3b}
\end{eqnarray}
where, it is easy to show that $C=(\partial B/\partial p_{1})-(\partial
\delta _{1}/ \partial p_{1}) A$, and that $D=-(\partial A/\partial
p_{1})-(\partial \delta _{1}/ \partial p_{1}) B$. The matrix element $%
P_{E_{1}E_{2}}$ is
\begin{equation}
P_{E_{1}E_{2}} \sim \frac{-iM}{\hbar \pi \sqrt{p_{1}p_{2}}}\left(
p_{1}+p_{2}\right) \tilde{A},  \label{comput3c}
\end{equation}
which according to the relations among matrix elements derived in the
previous subsection is related to $X_{E_{1}E_{2}}$ by Eq. (\ref{prop2}). The
remaining matrix element $(XP)_{E_{1}E_{2}}$, on the other hand, can be
computed from the element $X_{E_{1}E_{2}}^2$ according to Eq. (\ref{prop4})

\subsubsection{The integrals $A(p_{1},p_{2})$ and $B(p_{1},p_{2})$}

Thus, we are finally left with the computation of the integrals (\ref
{comput2a}) and (\ref{comput2b}). The integral $B\left( p_{1},p_{2}\right) $
of Eq. (\ref{comput2b}) is dominated by its upper limit $\bar{x}$
\begin{eqnarray}
B\left( p_{1},p_{2}\right)  &\sim &\frac{1}{2}\int^{\bar{x}}dx\;\cos \left[
\left( p_{1}-p_{2}\right) x/\hbar +\delta _{1}-\delta _{2}\right]   \nonumber
\\
\  &\sim &\frac{1}{2\left( p_{1}-p_{2}\right) }\sin \left[ \left(
p_{1}-p_{2}\right) \bar{x}/\hbar +\delta _{1}-\delta _{2}\right]   \nonumber
\\
\  &\rightarrow &\frac{\pi \hbar }{2}\delta \left( p_{1}-p_{2}\right) ,
\label{comput4a}
\end{eqnarray}
The integrals $A_{A,S}\left( p_{1},p_{2}\right) $ are more subtle. The
integral $A_{S}$ is clearly regular on the diagonal. Since we are interested
mostly on the singular behavior of the matrix elements, we can approximate $%
A_{S}\sim 0.$ On the other hand $A_{A}$ is exactly zero on the diagonal.
Close to the diagonal, the integral is dominated by the region where the
argument of the trigonometric function is small, and thereby the integrand
is non oscillatory. Estimating the upper limit of this region as $\bar{x}%
\sim \hbar \left( p_{1}-p_{2}\right) ^{-1}$, we get
\begin{equation}
A_{A}\sim \hbar ^{-1}\left( p_{1}-p_{2}\right) \bar{x}^{2}+\left( \delta
_{1}-\delta _{2}\right) \bar{x}=\hbar PV\left( \frac{1}{p_{1}-p_{2}}\right)
+...\,,  \label{comput4b}
\end{equation}
where the dots stand for regular terms. Actually, this argument would allow
us to introduce an undetermined coefficient in front of the principal value $%
PV$, but in the next section we show that $\hbar $ is the correct
coefficient, as follows from the canonical commutation relations.

Thus, we are now in the position to give the explicit expressions for the
singular parts of the matrix elements and write, finally, the quantum
transport equation in its explicit form.
This is done in detail in the next section. However,
there is an approximation we can use that drastically simplifies the
computations, and is discussed afterwards, in subsection \ref{phaseshift}.

\section{The quantum Kramers equation}

\label{ab}

In this Section we explicitly compute the quantum transport equation (\ref
{open1}) satisfied by the reduced Wigner function in the energy
representation.

\subsection{Matrix elements}

First, we need to compute the matrix elements described in section \ref
{computmatrix}. We begin with the matrix element $X_{E_{1}E_{2}}$ which
according to (\ref{comput3a}) and (\ref{comput4a})-(\ref{comput4b}) can be
written as:
\begin{equation}
X_{E_{1}E_{2}}=\frac{ M\hbar}{\sqrt{p_{1}p_{2}}}\left[ \frac{1}{\pi }\frac{%
\partial }{\partial p_{1}} PV\left(\frac{1}{p_{1}-p_{2}}\right) -\frac{%
\partial \delta _{1}}{\partial p_{1}}\delta \left( p_{1}-p_{2}\right)
+...\right].  \label{ab1}
\end{equation}
We go next to the matrix element $P_{E_{1}E_{2}}$, which from (\ref{comput3c}%
) and (\ref{comput4b}) can be written as,
\begin{equation}
P_{E_{1}E_{2}} =\frac{-iM}{\sqrt{p_{1}p_{2}}}\frac{1}{2\pi }\left(
p_{1}+p_{2}\right) PV\frac{1}{p_{1}-p_{2}}.  \label{ab2}
\end{equation}
These two operators $X$ and $P$ are connected through Eq. (\ref{prop2}). It
is easy to check that the two previous results satisfy this relation. Just
notice that from Eq. (\ref{norm2}) we can write $E_1-E_2=(p_1^2-p_2^2)/2M$
which together with Eq. (\ref{ab1}) for $X_{E_1 E_2}$ lead to $-i\hbar/M$
times the right hand side of Eq. (\ref{ab2}), that is
\[
\left( E_{1}-E_{2}\right) X_{E_{1}E_{2}}=-\frac{i\hbar }{M}P_{E_{1}E_{2}}.
\]

Another check of the previous results is the consistency with the canonical
commutation relations
\begin{equation}
\int dE\;\left(P_{E_1E}X_{EE_2}- X_{E_1E}P_{EE_2} \right) =-i\hbar \delta
\left( E_1-E_2\right) .  \label{ab3}
\end{equation}
This check requires a little more work. First it is convenient to change to
momentum variables and write, $\delta \left( E_{1}-E_{2}\right) =(M/\sqrt{%
p_{1}p_{2}})\delta \left( p_{1}-p_{2}\right). $ Then one needs to compute
the integral
\begin{equation}
I \equiv \hbar \int_{-\infty }^{\infty }dp\; PV\left(\frac{1}{p_{1}-p}%
\right) PV\left( \frac{1}{p-p_{2}}\right)=-\hbar \pi ^{2}\delta \left(
p_{1}-p_{2}\right),  \label{ab4}
\end{equation}
The evaluation of this integral is easily performed using the following
representation of the principal value
\[
PV\left( \frac{1}{p}\right) =\int \frac{d\xi }{2\pi\hbar }\;e^{ip\xi/\hbar
}\left( -i\pi \;\mathrm{sign}\left[ \xi \right] \right),
\]
which is easily proved by taking the Fourier transform of $PV (1/p)$. With
the result of Eq. (\ref{ab4}) it is straightforward to check that the
commutation relation (\ref{ab3}) is an identity within our approximation.
This consistency check is important because it can be used to fix to $\hbar$
the coefficient in front of the principal value of $A_A$ in the argument
leading to Eq. (\ref{comput4b}).

We can now move to the matrix elements for $X^2$. Having an expression for $%
X_{E_{1}E_{2}}$ in Eq. (\ref{ab1}) it is best to compute $X_{E_{1}E_{2}}^{2}
$ directly from the relation (\ref{prop3}) which leads to
\begin{eqnarray}
X_{E_{1}E_{2}}^{2} &=&\frac{M\hbar^2}{\sqrt{p_{1}p_{2}}}\left[ \frac{%
\partial ^{2}}{\partial p_{1}\partial p_{2}}\delta \left( p_{1}-p_{2}\right)
+\frac{1}{\pi }\left( \frac{\partial \delta _{1}}{\partial p_{1}}+\frac{%
\partial \delta _{2}}{\partial p_{2}}\right) \frac{\partial }{\partial p_{2}}
PV\left(\frac{1}{p_{1}-p_{2}}\right) \right.  \nonumber \\
&&\left. +\left( \frac{\partial \delta _{1}}{\partial p_{1}}\right)
^{2}\delta \left( p_{1}-p_{2}\right) +...\right],  \label{ab5}
\end{eqnarray}
where we have used the result (\ref{ab4}) and performed the $E$ integration
or, more precisely, the $p$ integration.

The matrix element $(XP)_{E_{1}E_{2}}=\int dE X_{E_{1}E} P_{EE_{2}}$ can be
analogously obtained from the expressions (\ref{ab1}) and (\ref{ab2}). The
result is
\begin{equation}
\left( XP\right) _{E_{1}E_{2}}= \frac{iM\hbar}{2 \sqrt{p_{1}p_{2}}}\left[
2p_{2}\frac{\partial }{\partial p_{1}}\delta \left( p_{1}-p_{2}\right) +%
\frac{1}{\pi }\frac{\partial \delta _{1}}{\partial p_{1}}\left(
p_{1}+p_{2}\right) PV\left(\frac{1}{p_{1}-p_{2}}\right) +...\right].
\label{ab6}
\end{equation}
A further consistency check of these expressions comes from the property (%
\ref{prop4}), which is satisfied within our approximation.

\subsection{The quantum transport equation}

Finally, we can write the quantum transport equation in the energy
representation (\ref{open5}) with the coefficient $Q$ given by (\ref{open6a}%
). The values of the dissipative and noise parts are given, respectively, by
(\ref{open8a}) and (\ref{open8b}), which can be directly computed using the
matrix elements deduced in the previous subsection. It is convenient to
define,
\begin{equation}
C_{E_1E_2}\left( t\right) =\frac 1{\sqrt{p_1p_2}}C_{p_1p_2}\left( t\right),
\label{ab7}
\end{equation}
and the result is the rather cumbersome expression (\ref{ab8}) given in
Appendix \ref{aaa3}. As explained there we can get a local approximation of
the quantum transport equation (\ref{ab8}), namely
\begin{eqnarray}
\frac{\partial C_{p_1p_2}}{\partial t}&=& \frac{-i}{2M\hbar}%
(p_1^2-p_2^2)C_{p_1p_2} +\frac {\gamma}{2}\left( \frac \partial {\partial
p_1}+\frac \partial {\partial p_2}\right)\left[ \left( p_1+p_2\right)
C_{p_1p_2}\right]  \nonumber \\
&& +\gamma M\sigma^2 \left[\left( \frac \partial {\partial p_1}+\frac
\partial {\partial p_2}\right) ^2 - \left( \frac{\partial \delta _1}{%
\partial p_1} -\frac{\partial \delta _2}{\partial p_2}\right) ^2 \right]
C_{p_1p_2}.  \label{ab9}
\end{eqnarray}

It is suggestive to give interpretations to the last three terms in this
quantum Kramers equation. The first, of course, is the dissipation term,
whereas the second and third are diffusion terms. The first involves the
dissipation coefficient, that defines a time scale $\tau_R\sim \gamma^{-1}$,
which is the relaxation time.

Before we go on with the interpretation of the different terms, it is
important to recall the meaning of the coefficients $C_{p_1 p_2}$, or $%
C_{E_1 E_2}$. First, we note that these coefficients are directly related to
the coefficients $C_E$ of the energy eigenfunctions which make the tunneling
state from the false vacuum in the isolated system, \textit{i.\;e.} when
there is no interaction to the environment. Thus, the coefficients $C_{E_1
E_2}$ clearly give the quantum correlations between wave functions of
different energies that make the tunneling system. These coefficients are
initially separable $C_{E_1 E_2}(0)= C_{E_1}(0) C_{E_2}^\ast(0)$. In the
isolated closed system its time evolution, as given by Eq. (\ref{ener6}), is
simply $C_{E_1 E_2}(t)=C_{E_1 E_2}(0)\exp[-i(E_1-E_2)t/\hbar]$, which means
that these correlations keep their amplitude in its dynamical evolution.

This is very different in the open quantum system. The negative last term in
Eq. (\ref{ab9}) has no effect when $E_{1}=E_{2}$, \textit{i.\;e.} for the
diagonal coefficients, but its effect is very important for the off-diagonal
coefficients. In fact, the amplitude of the off-diagonal coefficients
exponentially decays in time, on a time scale of the order of
\begin{equation}
\tau _{D}\sim \tau _{R}\left( \frac{\lambda _{B}}{l_{D}}\right) ^{2},
\label{ab12}
\end{equation}
where $\tau _{R}$ is the relaxation time, $\lambda _{B}=\hbar /(2\sigma
\sqrt{M})$ is a characteristic de Brolie wavelength (in the high temperature
case when $\sigma ^{2}=k_B T$ it corresponds to the thermal de Broglie
wavelength), and $l_{D}\sim \alpha ^{2}\hbar \sqrt{E_{0}+U_{\infty }}%
/(\varepsilon \sqrt{M})$ is a characteristic length of the problem with $%
\alpha $ a dimensionless parameter that measures the scale of the energy
differences of the off-diagonal coefficient, $E_{1}-E_{2}\sim \alpha
\varepsilon $; so it is of order 1 when the energy differences are of order $%
\varepsilon $. The time scale (\ref{ab12}) can be estimated by taking the
derivatives of the phase shifts $\delta _{i}$ ($i=1,2$) near the false
vacuum energy $E_{0}$, which is where the energy wavefunctions pile up.
Thus, the last term of equation (\ref{ab9}) destroys the quantum
correlations of the energy eigenfunctions. The time scale $\tau _{D}$ may be
considered as a decoherence time \cite{Zur91}, and thus the effect on
tunneling of this term may be associated to the effect of decoherence.

Another time scale in the problem is, of course, the tuneling time which
according to (\ref{tunn3}) and (\ref{norm13}) is given by $\tau_{\mathrm{tunn%
}}\sim \hbar/\varepsilon$. Its relation to $\tau_D$ is given by $\tau_D\sim
\tau_{\mathrm{tunn}} /(\alpha^4 D)$, where the dimensionless parameter $D$
is defined in (\ref{tunnel5}).

The last of the diffusion terms is the only one that survives in the phase
shift approximation which we introduce in the next subsection. This is
justified by the strong dependence of the phase shifts $\delta_i$, ($i=1,2$)
on the energy near the vacuum energy, see Eq. (\ref{norm15b}), which make
the derivatives of these functions very large near $E_0$. Note also that the
range of energies (and momenta) in Eq. (\ref{ab9}) is also limited to near $%
E_0$ as the coefficients $C_E$ that describe the tunneling state are peaked
there; cf. Eq. (\ref{vacu4}).

\subsection{The phase shift approximation}

\label{phaseshift}

The phase shift approximation is based on the observation made in Section
\ref{wkb} that the phase shifts are fast varying functions of energy near
resonance. This suggests: (a) we only keep the singular terms, and of these,
(b) only those which contain derivatives of the phase shifts. Under this
approximation we can go back to Eqs. (\ref{comput3a})-(\ref{comput3c}) to
write,
\begin{equation}
X_{E_1E_2}\sim \frac{-\hbar M}{p_1}\left( \frac{\partial \delta _1}{\partial
p_1}\right) \delta \left( p_1-p_2\right),  \label{phase1a}
\end{equation}
\begin{equation}
X_{E_1E_2}^2\sim \frac {\hbar^2 M}{\sqrt{p_1p_2}}\left( \frac{\partial
\delta _1}{\partial p_1}\right) ^2\delta \left( p_1-p_2\right),
\label{phase1b}
\end{equation}
moreover $P_{E_1E_2}\sim 0$, and $(XP)_{E_1E_2}\sim 0$. Now we see from (\ref
{open8a}) that $Q_{E_1E_2,E_1^{\prime }E_2^{\prime }}^{\left( D\right)}\sim
0 $, so that only the diffusion term $Q_{E_1E_2,E_1^{\prime }E_2^{\prime
}}^{\left( N\right)}$ matters, and this term is proportional to $\delta
\left( E_1-E_1^{\prime }\right) \delta \left( E_2-E_2^{\prime }\right)$ (we
change from variables $p_i$ to $E_i$ according to $p_{i }dp_{i }=M dE_{i}$).
This means that we are working in a weak dissipation limit. Finally, the
quantum transport equation (\ref{open5}) can be written as
\begin{equation}
\frac \partial {\partial t}C_{E_1E_2}\left( t\right) =-L\left[
E_1,E_2\right] C_{E_1E_2}\left( t\right),  \label{phase2}
\end{equation}
where
\[
L\left[ E_1,E_2\right] =\frac i\hbar \left( E_1-E_2\right) +\frac{\gamma
\sigma^2}{M}\left( p_1\frac{\partial \delta _1}{\partial E_1}-p_2\frac{%
\partial \delta _2}{\partial E_2}\right) ^2.
\]

This equation, of course, can also be obtained from Eq. (\ref{ab9}) in the
limit where only the phase shift terms of the environment are kept and we
return to the $E_{i}$ variables instead of the $p_{i}$. In the spirit of the
phase shift approximation, we shall replace $p_{i}$ ($i=1,2$) by their
values at resonance, whereby
\begin{equation}
L\left[ E_{1},E_{2}\right] =\frac{i}{\hbar }\left( E_{1}-E_{2}\right)
+2\gamma \sigma ^{2}(E_{0}+U_{\infty })\left( \frac{\partial \delta _{1}}{%
\partial E_{1}}-\frac{\partial \delta _{2}}{\partial E_{2}}\right) ^{2},
\label{phase3}
\end{equation}
where, from Eq. (\ref{norm15b}), the phase shifts derivatives are
\begin{equation}
\frac{\partial \delta }{\partial E}=\frac{-i}{2}\left( \frac{1}{E-E_{-}}-%
\frac{1}{E-E_{+}}\right) .  \label{phase4}
\end{equation}

\section{Tunneling in the open quantum system}

\label{tunnel}

We can now compute the tunneling rate from the false vacuum for our open
quantum system. Thus, let us assume that our particle at $t=0$ is trapped
into the well of the potential (\ref{potential}) in the false ground state
with the energy $E_{0}$, \textit{i.\;e.}
the ground state of the auxiliary potential
$U_{\mathrm{aux}}$ introduced in Section \ref{vacu}. We know from that
section that the wave function $\psi _{0}$ of this state can be expressed in
terms of the eigenfunctions $\psi _{E}$ by Eq. (\ref{vacu1a}) with the
coefficients $C_{E}$ given by Eq. (\ref{vacu4}). In terms of the reduced
Wigner function, which we may call $W_{0}(x,p)$, this state is easily
described in the energy representation (\ref{ener5}) by the coefficients $%
C_{E_{1}E_{2}}(0)=C_{E_{1}}(0)C_{E_{2}}^{*}(0)$, where $C_{E}(0)$ is just
given by Eq. (\ref{vacu4}). Because the dynamics of the quantum transport
equation is trivial in the energy representation (\ref{phase2}) the time
dependence of the coefficients $C_{E_{1}E_{2}}(t)$ is simply
\begin{equation}
C_{E_{1}E_{2}}\left( t\right) =\exp \left( -L\left[ E_{1},E_{2}\right]
t\right) C_{E_{1}E_{2}}\left( 0\right) ,  \label{tunnel1}
\end{equation}
so that, according to Eq. (\ref{ener5}), the Wigner function at any time is
\begin{equation}
W\left( x,p,t\right) =\int dE_{1}dE_{2}\;\exp \left( -L\left[
E_{1},E_{2}\right] t\right) C_{E_{1}}\left( 0\right) C_{E_{2}}\left(
0\right) W_{E_{1}E_{2}}\left( x,p\right) .  \label{tunnel2}
\end{equation}

{}From this we can compute, in particular, the probability of finding the
particle at the false vacuum at any time. In terms of the false vacuum
Wigner function and the Wigner function of the tunneling system we may
define that probability as
\begin{equation}
\rho^2(t)=2\pi\hbar \int dx\,dp\, W_0(x,p)W(x,p,t).  \label{tunnel2a}
\end{equation}
This equation is justified by observing that in the closed system of Section
\ref{sec2} where the state is described by the wave function $\psi$ of Eq. (%
\ref{tunn1}) and the false vacuum is described by the wave function $\psi_0$
of Eq. (\ref{vacu1a}), the square of the persistence amplitude (\ref{tunn2})
is given, in fact, by Eq. (\ref{tunnel2a}) when the definition of the Wigner
function , \textit{i.\;e.} Eq. (\ref{wigner1}), is used. For the open system
the quantum state is not described by a pure state and, in general, the
Wigner function $W(x,p,t)$ can be written as $W=\sum_i p_iW_i$ where $p_i$
is the probability of finding the system in the state $\phi_i$ and $W_i$ is
the Wigner function for the state $\phi_i$. The definition (\ref{tunnel2a})
leads in this case to $\rho^2(t)=\sum_i p_i |\langle \psi_0
|\phi_i\rangle|^2 $, which is indeed the probability of finding the system
in the state $\psi_0$. Eq. (\ref{tunnel2a}) when the energy representation (%
\ref{ener5}) is used becomes
\begin{equation}
\rho ^{2}\left( t\right) =\int dE_{1}dE_{2}\;\exp \left( -L\left[
E_{1},E_{2}\right] t\right) C_{E_{1}}^{2}\left( 0\right) C_{E_{2}}^{2}\left(
0\right).  \label{tunnel3a}
\end{equation}

To compute $\rho ^{2}\left( t\right) $ we shall use the stationary phase
approximation. The idea is that the integration paths for $E_{1}$ and $E_{2}$
may be deformed simultaneously in such a way that the integrand comes to be
dominated by Gaussian peaks. For late times it is enough to seek the
stationary points of $L\left[ E_{1},E_{2}\right] .$ In principle, we could
include $K_{E_{1}}^{2}$ and $K_{E_{2}}^{2}$ as fast varying components of
the integrand, but these functions are really fast varying in the vicinity
of $E_{-}$ and $E_{+},$ which, when $\gamma \neq 0$, are essential
singularities of the integrand and must be avoided. Note that when deforming
the path of integration, we should avoid regions where Re $L<0$. Then,
calling
\begin{equation}
E_{i}=E_{0}+\varepsilon r_{i}\ \ \ (i=1,2),\qquad D=\gamma \hbar \sigma ^{2}%
\frac{(E_{0}+U_{\infty })}{\varepsilon ^{3}},  \label{tunnel5}
\end{equation}
where $\varepsilon $ was introduced in Eq. (\ref{norm13}), and using Eq. (%
\ref{phase4}) for the phase shift derivatives we find that
\begin{equation}
L\left[ E_{1},E_{2}\right] t=\frac{t\varepsilon }{\hbar }\left[ i\left(
r_{1}-r_{2}\right) +2D\left( \frac{1}{1+r_{1}^{2}}-\frac{1}{1+r_{2}^{2}}%
\right) ^{2}\right] .  \label{tunnel3b}
\end{equation}
The stationary phase condition, $dL/dE_{1}=0$ and $dL/dE_{2}=0$, reads
\begin{equation}
f_{1}\left[ r_{1},r_{2}\right] \equiv i-8D\left( \frac{1}{1+r_{1}^{2}}-\frac{%
1}{1+r_{2}^{2}}\right) \frac{r_{1}}{\left( 1+r_{1}^{2}\right) ^{2}}=0,
\label{primera}
\end{equation}
and there is a similar equation $f_{2}\left[ r_{1},r_{2}\right] =0$, where $%
f_{2}\left[ r_{1},r_{2}\right] $ is obtained from $f_{1}\left[
r_{1},r_{2}\right] $, with the substitution of the multiplying factor $%
r_{1}/\left( 1+r_{1}^{2}\right) ^{2}$ by $r_{2}/\left( 1+r_{2}^{2}\right)
^{2}$. Observe that $f_{2}\left[ r_{1},r_{2}\right] =-\left( f_{1}\left[
r_{2}^{*},r_{1}^{*}\right] \right) ^{*}$, so a solution of Eq. (\ref{primera}%
) with $r_{2}=r_{1}^{*}$ is automatically also a solution of $f_{2}\left[
r_{1},r_{2}\right] =0$. We shall seek stationary points of this kind. It is
clear that for a solution of this kind, $r_{1}/\left( 1+r_{1}^{2}\right) ^{2}
$ must be real. So, writing $r_{1}=\xi -i\eta $, we must have $4\xi
^{2}\left( 1+\xi ^{2}\right) =\left( 1+\xi ^{2}-\eta ^{2}\right) ^{2}$, and
solving for\ $\xi ^{2}$ we find
\begin{equation}
3\xi ^{2}=\left[ \left( 2-\eta ^{2}\right) ^{2}+3\eta ^{4}\right]
^{1/2}-1-\eta ^{2}.  \label{cuarta}
\end{equation}
We notice that for each value of $\eta $ there will be two possible
solutions for $\xi $. The path of integration must go through both of them.
Substituting, $r_{1}=\xi -i\eta $, into the complex stationary phase
equation (\ref{primera}) we obtain two equations for $\xi $ and $\eta $. The
real part of Eq. (\ref{primera}) leads to the previous Eq. (\ref{cuarta}),
which is independent of $D$, and the imaginary part leads to
\begin{equation}
32D\xi ^{2}\eta \left[ \left( 1+\xi ^{2}-\eta ^{2}\right) ^{2}+4\eta
^{2}\left( 1-\eta ^{2}\right) \right] =\left[ \left( 1+\xi ^{2}-\eta
^{2}\right) ^{2}+4\xi ^{2}\eta ^{2}\right] ^{3}.  \label{tercera}
\end{equation}
For each value of $D$, we are interested in the solutions with the lowest
possible positive value of $\eta $.

Finally, the contribution of each saddle point to the integral will be
\begin{equation}
\rho _{\pm }^{2}(t)=\Delta _{\pm }e^{-tL\left[ \pm \xi ,\eta \right] },
\label{quinta}
\end{equation}
where
\[
L\left[ \pm \xi ,\eta \right] =\frac{2\varepsilon\eta}{\hbar} \left(
1-16D\xi ^{2}\eta \left[ \left( 1+\xi ^{2}-\eta ^{2}\right) ^{2} +4\xi
^{2}\eta ^{2}\right] ^{-2} \right),
\]
and where the prefactor $\Delta _{\pm }$ depends on the second derivatives
of $L$. Comparing to the persistence probability $\rho _{\mathrm{closed}%
}^{2}\sim \exp \left( -2t\varepsilon /\hbar \right) $ for the isolated
closed quantum system, which follows from the persistence amplitude Eq. (\ref
{tunn3}) and the definition of $\varepsilon$ given in Eq. (\ref{norm13}), we
conclude that the ratio $R=\Gamma_{open}/\Gamma_{closed}$ of the tunneling
rates between the open, $\Gamma_{open}=L$, and closed, $\Gamma_{closed}=2%
\varepsilon/\hbar$, systems is
\begin{equation}
R=\eta \left( 1-16D\xi ^{2}\eta \left[ \left( 1+\xi ^{2}-\eta ^{2}\right)
^{2}+4\xi ^{2}\eta ^{2}\right] ^{-2}\right) ,  \label{sexta}
\end{equation}
where the parametres $\xi$ and $\eta$ are solutions of the algebraic
equations (\ref{cuarta}) and (\ref{tercera}) with the lowest possible
positive value of $\eta$.

\begin{figure}[tbp]
\includegraphics[height=8cm]{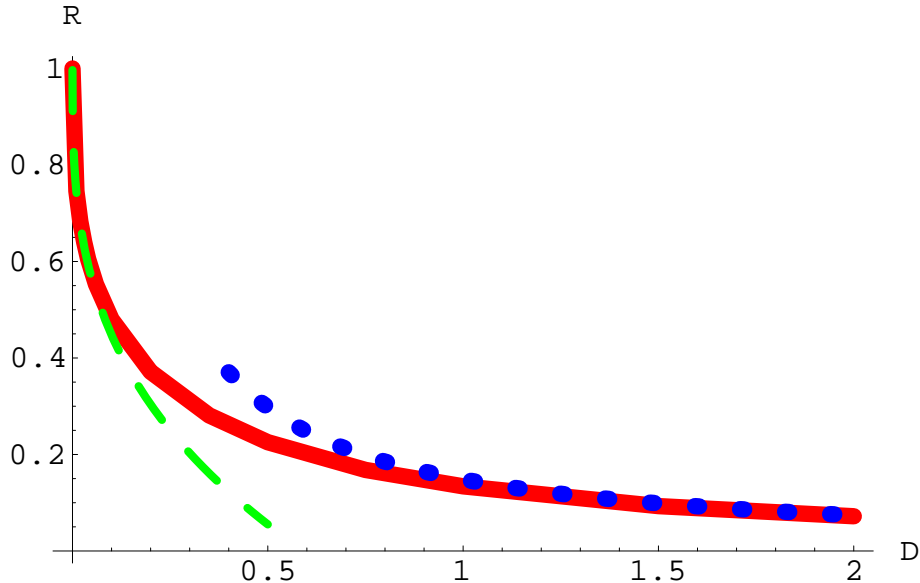}
\caption{The ratio $R$ between the tunneling rates for the open system and
the closed system as a function of the dimensionless parameter $D$ (full
line). We also plot the analytic approximations for low and large values of $%
D$ (dashed and dotted lines, respectively).}
\end{figure}

In Fig. 2 we plot $R$ as a function of $D$ from a numerical solution of Eqs.
(\ref{tercera}) and (\ref{cuarta}) (full line); we see that $R<1$ for all
values of $D,$ with tunneling being strongly suppressed when $D$ is large.
In the asymptotic limits $D\rightarrow 0$ and $D\rightarrow \infty $ it is
possible to obtain analytical approximations. In the former, $\eta =1-\delta
,$ $\xi =\pm \delta ,$ where $\delta =\left( D/2\right) ^{1/3}.$ In this
limit
\begin{equation}
R_{D\rightarrow 0}\sim 1-\frac{3}{2}\left( \frac{D}{2}\right) ^{1/3},
\label{sexta.1}
\end{equation}
which is also plotted in Fig. 2 (dashed line). In the opposite limit, $%
D\rightarrow \infty ,$ we have $\eta =8/\left( 27D\right) ,$ $\xi =\pm 1/%
\sqrt{3}$ and
\begin{equation}
R_{D\rightarrow \infty }\sim \frac{4}{27D},  \label{sexta.2}
\end{equation}
which is plotted in Fig. 2 (dotted line).

Observe that if we included the factors of $K_{E_{i}}^{2}$ ($i=1,2$) into
the fast varying part of the integral, then the saddle points shift by an
amount $\sim \hbar /(\varepsilon t)$; therefore these results are reliable
when $(\varepsilon t/\hbar )D^{1/3}\gg 1.$ On the other hand, when $%
\varepsilon t/\hbar $ is extremely large, the energy integrals are dominated
by the contribution from the lower limit $-U_{\infty }$ and the decay rate
turns to a power law.

\section{Comparing with experiment}

\label{experiment}

In this section we confront with experimental results. One of the
experiments in which macroscopic quantum tunneling has been observed in more
detail is a single Josephson junction between two superconducting electrodes
biased by an external current. The macroscopic variable in this case is the
phase difference $\varphi$ of the Cooper pair wave function across the
junction \cite{CalLeg83b}. A macroscopic system always interacts with an
environment and a physical Josephson junction is generally described by an
ideal one shunted by a resistance and a capacitance which phenomenologically
account for the effects of the environment.

The basic equations relate the (gauge invariant) phase difference
$\varphi, $  the current $I$ and the voltage $V$ across an ideal
junction \cite {Tin96,MarDevCla87}. A zero voltage supercurrent
$I=I_{c}\sin \varphi, $ should flow between two superconducting
electrodes separated by a thin insulating barrier, where $I_{c}$
is the critical current, the maximum supercurrent the junction can
support. If a voltage difference $V$ is
maintained across the junction the phase difference evolves according to $%
2eV=\hbar d\varphi /dt$, where $2e$ is the charge of a Cooper
pair, and $\hbar /(2e)$ is the flux quantum. It is convenient to
introduce the characteristic energy scale $E_{J}=\hbar I_{c}/2e$;
note that the work necessary to raise the phase difference across
the ideal junction from $0$ to $\varphi $ is $\left( -E_{J}\right)
\cos \varphi $. A real junction is modelled as an ideal one in
parallel with an ordinary resistance $R_{\Omega } $, which build
in dissipation in the finite voltage regime, and an ordinary
capacitance $C$, which reflects the geometric shunting capacitance
between the two superconducting electrodes. The so-called bias
current flowing through the device is $I=I_{c}\sin \varphi
+V/R_{\Omega }+CdV/dt$, and substituting the previous relationship
between $V$ and $d\varphi /dt$, this equation becomes a
differential equation for the phase difference of the Cooper pair
$\varphi $:
\begin{equation}
M\left( \frac{d^{2}\varphi }{dt^{2}}+\gamma \frac{d\varphi }{dt}\right) +%
\tilde{U}^{\prime }\left( \varphi \right) =0,  \label{7.1}
\end{equation}
which has the form of the equation of motion for a particle in a
one-dimensional potential with friction. We have introduced the ``mass'' $%
M=\hbar ^{2}C/(2e)^{2}$, the friction coefficient $\gamma =1/(R_{\Omega }C)$%
, and the ``potential''
\begin{equation}
\tilde{U}\left( \varphi \right) =\left( -E_{J}\right) \left( \cos \varphi
+s\varphi \right) ,  \label{7.2}
\end{equation}
where $s\equiv I/I_{c}$. This is the so-called ``tilted
washboard'' model. Note that when $s=1$ the local minima of the
tilted cosine become inflection points, and that no classical stable
equilibrium points exist when $s\geq 1$. We will be interested in
the case in which $s$ is smaller but close to $1$, that is when
the external biased current $I$ is slightly less than the critical
current $I_{c}$.

In this case, the potential may be approximated by a cubic potential in the
neighborhood of any stable stationary point. Let us consider the stable
stationary point closest to $\varphi =0.$ The stationarity condition is $%
\sin \varphi =s$ and the stability condition is $\cos \varphi >0$. We see
that there must be a solution $\varphi _{0}\leq \pi /2$, thus let us write $%
\varphi _{0}=\frac{\pi }{2}-\kappa $, where $\kappa \sim \sqrt{2\left(
1-s\right) }\sim \sqrt{1-s^{2}}$. We henceforth introduce a new variable $%
x=\varphi -\varphi _{0}$ and the shifted potential
\begin{equation}
U\left( x\right) =\tilde{U}\left( \varphi _{0}+x\right) -\tilde{U}\left(
\varphi _{0}\right) ,  \label{7.3}
\end{equation}
which leads to the same equation (\ref{7.1}) than the potential (\ref{7.2}).
In the new variable, the stable stationary point lies at $x=0.$ The closest
unstable stationary point, the maximum of the potential lies at $%
x_{s}=2\kappa $, and following the notation of Sec. \ref{sec2}, we have that
the height of the potential barrier has an energy $\varepsilon _{s}\equiv
U\left( x_{s}\right) $, given by
\begin{equation}
\varepsilon _{s}\sim \frac{2}{3}E_{J}\left[ 2\left( 1-s\right)
\right] ^{3/2}\sim \frac{2}{3}E_{J}\left( 1-s^{2}\right) ^{3/2}.
\label{7.3a}
\end{equation}
Finally, the next root of the potential is $x_{exit},$ where $\tilde{U}%
\left( \varphi _{0}+x_{exit}\right) =\tilde{U}\left( \varphi _{0}\right) $,
which can be approximately written as $sx_{exit}-\sin (x_{exit}-\kappa )\sim
\kappa .$ Observe that $x_{exit}\sim 3\kappa \sim \left( 3/2\right) x_{s},$
which is similar to what happens with the cubic potential of Sec. \ref{sec2}%
. Note that when $s$ is nearly $1$ the height of the potential barrier $%
\varepsilon _{s}$ is much less than the potential difference between
adjacent wells and the potential can be approximated by a cubic potential.
For $\left| x\right| $ no much larger than $x_{exit},$ we may approximate $%
U\left( x\right) \sim \varepsilon
_{s}(x/x_{s})^{2}(1-x/x_{exit})/(1-x_{s}/x_{exit})$, from where we may
define the frequency $\Omega _{0}$ of small oscillations around $x=0$: $%
M\Omega _{0}^{2}=U^{\prime \prime }(0)$, which gives
\begin{equation}
\Omega _{0}^{2}\sim \frac{3\varepsilon _{s}}{2\left( 1-s^{2}\right) M}\sim
\frac{4e^{2}E_{J}}{\hbar ^{2}C}\left( 1-s^{2}\right) ^{1/2}\equiv \omega
_{p}^{2},  \label{7.4}
\end{equation}
where $\omega _{p}$ is the ``plasma frequency'' of the junction.
We may introduce $\omega _{p0}\equiv (4e^{2}E_{J}/\hbar
^{2}C)^{1/2}=(2eI_{c}/\hbar C)^{1/2}.$

In the literature there are several reported observations of
tunneling in this or more complex set-ups \cite
{DevMarCla85,MarDevCla87,CleMarCla88,WalEtAl03,FulDun74,SchEtAl85,%
SilLieGra88,Sil92,LiEtAl02,MonMygRiv02,MonMygRiv03,FisWalUst03}.
From this
wealth of data we have chosen to focus on comparison against the
experiments reported in Ref. \cite{CleMarCla88}. In that paper,
tunneling was observed under a fixed bias current, as opposed to a
time-dependent one \cite {FulDun74}. The fixed current environment
is closest to the ideal situation to which our model applies.

A critical input in comparing theory to observations is the value of
the critical current for the junction.
In Ref. \cite{CleMarCla88} the critical
current is not derived from the tunneling rate itself, as advocated
in Ref. \cite{GraOlsWei86}, but extracted
from an independent set of measurements
at high temperature. We therefore find that the comparison of theory
and experiment may be done in two ways: it is possible to derive the
escape temperature from the independently measured critical current,
as it is done in Ref. \cite{CleMarCla88}, or
else it is possible to induce the value
of the critical current from the observed rate. This value may then be
compared to the one obtained in high temperature determinations.
In the following we present both analysis.

\subsection{Tunneling rates for the open system}

Let us first compare our theoretical model against the
value for the tunneling amplitude (extrapolated up to) at $T=0,$
as reported in Ref. \cite {CleMarCla88}. The relevant values for
the bias current $I$, the critical current $I_{c}$, its ratio $s$,
the self-capacitance $C$, and shunt resistance $R_{\Omega }$ are:
\begin{equation}
I\sim 24.710\;\mu \mathrm{A},\ I_{c}\sim 24.873\;\mu \mathrm{A},\
s\sim 0.9934,\ C\sim 4.28\;\mathrm{pF},\ R_{\Omega }\sim
9.3\;\mathrm{\Omega }. \label{7.5}
\end{equation}
These parameters were measured in the classical limit. In previous similar
experiments on current-biased junctions but with low dissipation \cite
{DevMarCla85,MarDevCla87} it was shown that at low temperature the tunneling
rate became almost independent of temperature.

Since $\hbar =1.054572\times 10^{-34}\;\mathrm{Js}$ and
$e=1.602176\times 10^{-19}\;\mathrm{C}$ we have, in the
conditions of Ref. \cite{CleMarCla88}, $E_{J}\sim 8.185\times
10^{-21}\;\mathrm{J}$, which using Boltzmann's constant
$k_{B}=1.380650\times 10^{-23}\;\mathrm{JK}^{-1},$ may be
converted into a
temperature $E_{J}/k_{B}\sim 592.9\;\mathrm{K}.$ Therefore
the height of the barrier $%
\varepsilon _{s}$, $\omega _{p0}$, the plasma frequency $\Omega _{0}$, the
zero point energy $\varepsilon _{0}=\frac{1}{2}\hbar \Omega _{0}$, and the
friction coefficient $\gamma $ in our model are:
\begin{eqnarray}
&&\varepsilon _{s}/k_{B} \sim 589.74\;\mathrm{mK},\ \omega
_{p0}\sim 132.88\times 10^{9}\;\mathrm{s}^{-1},\ \Omega _{0}\sim
44.918\times 10^{9}\;\mathrm{s}^{-1},\   \nonumber \\
&&\varepsilon _{0}/k_{B} \sim 171.55\;\mathrm{mK},\ \gamma \sim
25.123\times 10^{9}\;\mathrm{s}^{-1}.  \label{7.6}
\end{eqnarray}
We observe that the zero point energy is lower than the barrier, although
not much lower. In fact, there is only one trapped excited state.

It is often convenient to give tunneling rates in terms of an
effective escape temperature $T_{esc}$. This escape temperature is defined
\cite{MarDevCla87} from a given tunneling rate, $\Gamma $, by the equation
\begin{equation}
\Gamma \equiv \frac{1}{2\tau }e^{-\varepsilon _{s}/k_{B}T_{esc}}.
\label{tesc}
\end{equation}
For the closed system, either the WKB approximation or the
instanton method yield $T_{esc}\sim 71\;\mathrm{mK,}$ with a
barrier penetrability $\Lambda$, defined by
$\Gamma_{closed}=(1/2\tau)\exp(-\Lambda)$ (which gives
$\Lambda=(2/\hbar)S_0$ in the WKB approximation),
of $\Lambda =8.459$; see Eq. (\ref{D10})
in Appendix D.

For the open system, the tunneling rate in our model
may be expressed in terms of $R$,
introduced in Eq. (\ref{sexta}) as the ratio between the open and closed
systems rates, namely
\begin{equation}
\Gamma _{open}=R\,\Gamma _{closed}=\frac{R}{2\tau }e^{-\Lambda }.
\label{7.10}
\end{equation}
The relevant parameter in $R$ is $D$ which was introduced in
(\ref {tunnel5}), where at zero temperature $\sigma ^{2}\sim
\varepsilon _{0}$ and also $\varepsilon _{0}\sim E_{0}$. Thus we
have $D=16\pi ^{3}(\gamma /\Omega _{0})(1+U_{\infty }/E_{0})\exp
(3\Lambda )$. With the numerical values given in
(\ref{7.5})-(\ref{7.6}), $D$ is very large and we are in the limit
of Eq. (\ref{sexta.2}), namely $R\sim 4/(27D)$. Comparing Eq.
(\ref{7.10}) with Eq. (\ref{tesc}), and assuming $U_{\infty
}/E_{0}<1$, we may find our predicted effective escape temperature from
$\varepsilon _{s}/(k_{B}T_{esc})\sim \Lambda +\ln (27D/4)\sim
4\Lambda +8.\,827$, which leads to $T_{esc}\sim
14.\,255\;\mathrm{mK}.$

The experimental result \cite{CleMarCla88} when extrapolated to the zero
temperature limit is, in terms of the effective escape temperature, $%
T_{esc,exp}\sim 45\;\mathrm{mK}$. The experimental
result is in good agreement with the
instanton prediction of Caldeira and Leggett for weak dissipation at zero
temperature \cite{CalLeg83b}; see also \cite{GraWei84,HanTalBor90,Wei93}.
Our prediction goes in the sense that
dissipation also suppresses tunneling but the predicted value
for the effective escape temperature is lower than the
observed result. We recall that our prediction is based on Eq. (\ref{phase2})
which was derived under an approximation that
fully neglects any activation that would increase
the effective escape temperature.
A better approximation based on Eq. (\ref{ab9}) would require
a numerical solution, but our starting
phenomenological equation (\ref{open1})
is too crude to expect a quantitative agreement
with the experimental result at very low temperatures at the level of accuracy
of the experiment.

\subsection{The critical current}

The value of the critical current is a crucial input in the
calculation of the tunneling rate \cite{GraOlsWei86}.
We are primarily interested in the tunneling rate under fixed
bias current. However, the critical current was not determined
under these conditions, but extracted from the average of
measurements of tunneling rates under a ramped bias current
performed at several different temperatures. For this reason, it
is meaningful to contrast theory and experiment in a different way
than in the previous subsection, namely, instead of computing the
tunneling rate from the given values of the critical and bias
currents, we may instead compute the critical current from the
given bias current and tunneling rate. We may then see if the
critical current obtained this way is consistent from that
obtained from the ramped current measurements also reported in
Ref. \cite{CleMarCla88}.

Recalling that the relation between the critical and bias currents
is $I_{c}=I/s$, where we will now assume that $s$ is so far
unknown but close to unity, we may write $E_{J}=\hbar I/(2es)$.
Then, from (\ref{7.3a}) we have that the height of the potential
barrier in terms of the parameter $s$ is, $\varepsilon _{s}\sim
(\hbar I/3e)s^{-1}(1-s^2)^{3/2}$, and from (\ref{7.4}) the frequency
of small oscillations is $\Omega _{0}\sim (2eI/\hbar C)^{1/2}
s^{-1/2} (1-s^2)^{1/4}.$ We shall assume that the ground
state energy is close to the zero point energy
$E_{0}\sim \varepsilon_0=\hbar \Omega _{0}/2$.
The ratio of the ground state energy to the height
of the barrier is thus, in terms of $s$,
\begin{equation}
\frac{E_{0}}{\varepsilon _{s}}\sim \frac{\bar\rho \,s^{1/2}}{(
1-s^{2}) ^{5/4}}, \label{7.11}
\end{equation}
where $ \bar\rho \equiv (9e^{3}/2\hbar CI)^{1/2}\sim 1.288\times
10^{-3}.$

As we discuss in Appendix D, due to the form of the potential the
action under the barrier, $S(x_{out},x_R)$, for the energy $E$ is equal
to the action inside the barrier, $S(x_R,x_L)$, for an energy
$E_{ref}=\varepsilon _{s}-E.$ Writing as in Eq. (\ref{D4}) $
E_{ref}=2\varepsilon _{s}\zeta \left( k_{ref}\right) $, where
$\zeta(k)$ is defined in Eq. (\ref{D7}), we get $ (1-s^{2})
^{5/4}s^{-1/2}=\bar\rho [1-2\zeta \left( k_{ref}\right)]^{-1/2} ,$
which relates the parameter $s$ with $k_{ref}$. The corresponding
penetrability, defined here as $\Lambda=(2/\hbar)S(x_{out},x_R)$, is
\begin{equation}
\Lambda =\frac{27}{8}\frac{F\left( k_{ref}\right) }{1-2\zeta
\left( k_{ref}\right) }, \label{7.12}
\end{equation}
where $F\left( k_{ref}\right) $ is defined in Eq. (\ref{D9}). The
tunneling rate according to Eq. (\ref{7.10}) is then
\begin{equation}
\Gamma_{open} =\frac{\Omega _{0}^{2}}{2\pi \gamma }\frac{4}{27\times
16\times \pi ^{3}}e^{-4\Lambda }. \label{7.13}
\end{equation}
We must compare this with the experimental tunneling rate which
may be computed from Eq. (\ref{tesc}) with the effective escape
temperature $T_{esc,exp}\sim 45\;\mathrm{mK}$, and with the
barrier hight $\varepsilon_s$ computed with a critical current
$I_{c}\sim 24.873\;\mu \mathrm{A}$. Finally, equating the
tunneling rate (\ref{7.13}) with the experimental one, using the
relevant numerical values given previously, we get
\begin{equation}
\frac{4F\left( k_{ref}\right) }{1-2\zeta \left( k_{ref}\right) }-\frac{2}{5}%
\ln \left[ \frac{1}{1-2\zeta \left( k_{ref}\right) }\right] =5.075+\frac{4%
}{5}\ln \left( \frac{1}{s}\right) . \label{eqn3}
\end{equation}
This equation must be solved simultaneously with the previous
equation relating $s$ with $k_{ref}$. The solution is $k_{ref}\sim
0.1162$ and $s\sim 0.9968.$ Thus the critical current is
\begin{equation}
I_{c}\sim 24.789\;\mu \mathrm{A}. \label{7.14}
\end{equation}
Observe that our model neglects any contribution that leads to
activation, and therefore underestimates the tunneling
rate \cite{MarGra88}. In matching against experiment, this effect
is compensated by lowering the predicted critical current.
Therefore this result must be regarded as a lower bound. With this
in mind, the agreement with the value of $I_{c}\sim 24.873\;\mu
\mathrm{A}$ extracted from the ramped current measurements \cite
{CleMarCla88} is satisfactory.

\section{Conclusions}

\label{conclusions}

To conclude, let us briefly summarize our findings. We have computed
analytically the effect of decoherence on quantum tunneling in a model
representing a quantum particle, trapped in a local minimum of a potential,
which is coupled to an environment characterized by a dissipative and a
diffusion parameter. We have used the master equation for the reduced
Wigner function, or quantum transport equation, that describes this open
quantum system.

Our computational method involves the introduction of an energy
representation of the reduced Wigner function which is based on the energy
eigenfunctions of the isolated system. The dynamical equation in this
representation, Eq. (\ref{open5}), is an equation for some coefficients that
describe the quantum correlations between eigenfunctions of different
energies. The equation can be explicitly written in a local approximation,
the quantum transport equation (\ref{ab9}), that captures the essential
physics of the problem.

In our problem, where the particle is initially trapped in the false vacuum,
the transport equation is dominated by a term that destroys the quantum
correlations of the eigenfunctions and is, thus, responsible for
decoherence. The strength of this term is characterized by the dimensionless
parameter $D$, defined in Eq. (\ref{tunnel5}), which is directly
proportional to the energy difference between the false and true vacuum. It
does not seem surprising that decoherence suppresses tunneling, as it
destroys the fine tuning among the energy eigenfunctions that makes
tunneling possible in the isolated closed system. The analytic result for
the ratio of the tunneling rates between the open and isolated closed
systems is given by Eq. (\ref{sexta}). This ratio decreases, implying more
suppression, with increasing $D$.

Our model is based in the phenomenological transport
equation (\ref{open1}) for the
reduced Wigner function, which is a toy model at zero temperature.
Besides this assumption, in the paper we work at two
different levels of approximation. We first derive Eq. (\ref{ab9}), which still
retains the leading effects of noise, dissipation and decoherence. To be
able to isolate the effects of decoherence, we then proceed to derive the
simpler Eq. (\ref{phase2}). The actual predictions for tunneling rates are obtained
from this later equation, which in practice means
to pick up the environment terms that give the most decoherence, and presumably
the most tunneling suppression. This
means that we have neglected from the remaining terms any
activation that would increase the effective escape temperature.
Therefore the results from our model, derived from Eq. (\ref{phase2}), must be
regarded as a lower bound on the actual rates. This is consistent with our
goal, which is not to provide an alternative to the instanton calculations
in equilibrium, but to propose a starting point for a real time formulation
of nonequilibrium macroscopic quantum tunneling. When seen under this light,
it is encouraging that the detailed matching against experimental results
shows that our model not only captures the main effect, since
indeed the measured rate is substantially lower than the quantum
prediction for the closed system, but also yields a suitable estimate of the
critical current from the given tunneling rate.
One might try to improve on this prediction by numerically solving
the quantum Kramers equation (\ref{ab9}), on which we expect to report in a
separate publication. The extent of the discrepancy can be seen
also as a check on the validity of the phenomenological terms
introduced into the master equation for the reduced Wigner function at
low temperature.

One should note that our approximations are valid when the initial state is
the false vacuum. The results might differ, even qualitatively, when more
general initial conditions are assumed and the terms that we have neglected
in the quantum transport equation become relevant. It may not be possible in
such a case to solve analytically the quantum transport equation. Yet, we
should emphasize that this equation in the energy representation is much
simpler than in the standard phase space representation.

\begin{acknowledgments}

We are grateful to Daniel Arteaga, Bei-Lok Hu, Fernando Lombardo,
Diana Monteoliva, Renaud Parentani, Ray Rivers and Albert Roura
for interesting discussions and suggestions. This work has been
partially supported by the MEC Research Projects Nos. FPA2001-3598
and FPA2004-04582 and by Fundaci\'{o}n Antorchas. E.~C.\
acknowledges support from Universidad de Buenos Aires, CONICET,
Fundaci\'{o}n Antorchas and ANPCYT through grant 03-05229.

\end{acknowledgments}
\appendix

\section{WKB solution}

\label{aaa}

In this Appendix we solve the WKB problem posed in section \ref{wkb}. The
starting point are Eqs. (\ref{wkb1}), (\ref{wkb2}) and (\ref{wkb3}) with the
cubic potential of Eq. (\ref{potential}), we have to match the WKB solutions
in the different regions across the potential function.

\subsection{Matching from forbidden to allowed regions}

Let $x_{0}$ be a classical turning point $U\left( x_{0}\right) =E,$ and let $%
U^{\prime }\left( x_{0}\right) <0.$ Then to the left of $x_{0}$ we have a
forbidden region, the two corresponding independent WKB solutions of the
Schr\"{o}dinger equation (\ref{wkb1}) are
\begin{equation}
F_{\pm }\left( x_{0},x\right) =\frac{e^{\pm S\left( x_{0},x\right) /\hbar }}{%
\sqrt{2p\left( x\right) /\hbar }},  \label{wkb4}
\end{equation}
whereas to the right of $x_{0}$ the two independent solutions are
\begin{equation}
G_{\pm }\left( x,x_{0}\right) =\frac{e^{\pm iS\left( x,x_{0}\right) /\hbar }%
}{\sqrt{2p\left( x\right) /\hbar }},  \label{wkb5}
\end{equation}
and we wish to find the corresponding matching conditions. For $x\rightarrow
x_{0}^{-},$ we can Taylor expand the potential around $x_{0}$ and write $%
p\left( x\right) =\kappa \left( x_{0}-x\right) ^{1/2}$ and $S\left(
x_{0},x\right) =\frac{2}{3}\kappa \left( x_{0}-x\right) ^{3/2}$, where we
have introduced $\kappa =\sqrt{2M\left| U^{\prime }\left( x_{0}\right)
\right| }$. Similarly for $x\rightarrow x_{0}^{+},$ we have $p\left(
x\right) =\kappa \left( x-x_{0}\right) ^{1/2}$ and $S\left( x,x_{0}\right) =%
\frac{2}{3}\kappa \left( x-x_{0}\right) ^{3/2}.$

If we write $x-x_{0}=e^{i\pi }\left( x_{0}-x\right) $ then $iS\left(
x,x_{0}\right) =S\left( x_{0},x\right) $ and it would seem that simple
analytical continuation yields $G_{+}\left( x,x_{0}\right) \rightarrow
e^{-i\pi /4}F_{+}\left( x_{0},x\right) $. However, this is impossible,
recall that if we define the flux $J= -i (\psi ^{\ast }\partial_x \psi -\psi
\partial_x \psi ^{\ast }) $ then the Schr\"odinger equation implies flux
conservation $\partial_x J=0$. Now $G_{+}\left( x,x_{0}\right) $ has $J=1$
and therefore it cannot turn into $F_{+}\left( x_{0},x\right) ,$ which is
real, and has $J=0$. Thus, we try instead
\begin{equation}
G_{+}\left( x,x_{0}\right) \rightarrow e^{-i\pi /4}F_{+}\left(
x_{0},x\right) +\beta F_{-}\left( x_{0},x\right),
\end{equation}
and imposing flux conservation we obtain $\beta=(1/2)\exp(i\pi /4)$. We
therefore find the matching conditions
\begin{equation}
e^{\mp i\pi /4}F_{+}\left( x_{0},x\right) +\frac{1}{2}e^{\pm i\pi
/4}F_{-}\left( x_{0},x\right) \rightarrow G_{\pm}\left( x,x_{0}\right),
\label{wkb6}
\end{equation}
{}from were we finally obtain, using Eq. (\ref{wkb5}),
\begin{eqnarray}
F_{+}\left( x_{0},x\right) \rightarrow \frac{1}{\sqrt{2p\left(
x\right)/\hbar }} \cos \left( \frac{1}{\hbar}S\left( x,x_{0}\right) +\frac{%
\pi }{4}\right),  \label{wkb7}
\end{eqnarray}
and
\begin{eqnarray}
F_{-}\left( x_{0},x\right) \rightarrow \frac{2}{\sqrt{2p\left(
x\right)/\hbar }} \sin \left( \frac{1}{\hbar}S\left( x,x_{0}\right) +\frac{%
\pi }{4}\right).  \label{wkb8}
\end{eqnarray}

\subsection{Matching from allowed to forbidden regions}

Now consider the case when $U^{\prime }\left( x_{0}\right) >0.$ To the left
of $x_{0},$ we have an allowed region and the solutions are oscillatory $%
G_{\pm }\left( x_{0},x\right) $, to the right of the turning point we have a
forbidden region and the solutions are a linear combination of (\ref{wkb4}).
By exactly the same procedure of the previous section, after imposing flux
conservation across $x_{0}$ we obtain:
\begin{equation}
G_{\mp }\left( x_{0},x\right) \rightarrow e^{\pm i\pi /4}F_{+}\left(
x,x_{0}\right) +\frac{1}{2}e^{\mp i\pi /4}F_{-}\left( x,x_{0}\right) .
\label{wkb9}
\end{equation}
Note from these equations that the solution that matches a decreasing
exponential is
\begin{equation}
\frac{1}{\sqrt{2p\left( x\right) /\hbar }}\sin \left( \frac{1}{\hbar }%
S\left( x_{0},x\right) +\frac{\pi }{4}\right) \rightarrow \frac{1}{2}%
F_{-}\left( x,x_{0}\right) .
\end{equation}

\subsection{WKB solution for $0<E<\varepsilon _{s}$}

We can now put all this together to write the energy eigenfunctions for our
cubic potential (\ref{potential}) for energies in the range $0<E<\varepsilon
_{s}$. There are three classical turning points in this case $%
x_{L}<x_{R}<x_{out}.$ To the left of $x_{L}$ we have a forbidden zone
extending to $-\infty ,$ so we have
\begin{equation}
\psi _{E}\left( x\right) \sim K_{E}F_{-}\left( x_{L},x\right) ;\qquad
x<x_{L},
\end{equation}
where $K_E$ is a normalization constant to be determined latter. To the
right of $x_{L}$ we have from (\ref{wkb8})
\begin{equation}
\psi _{E}\left( x\right)\sim \frac{2K_{E}}{\sqrt{2p\left( x\right)/\hbar }}%
\sin \left( \frac{1}{\hbar}S\left( x,x_{L}\right) +\frac{\pi }{4}\right),
\end{equation}
which after using the definition (\ref{wkb3}) can be rewritten in the region
$x_{L}<x<x_{R}$ as
\begin{equation}
\psi _{E}\left( x\right) \sim K_{E}\left( e^{i\left( S\left(
x_{R},x_{L}\right)/\hbar -\pi /4\right) }G_{-}\left( x_{R},x\right)
+e^{-i\left( S\left( x_{R},x_{L}\right)/\hbar -\pi /4\right) }G_{+}\left(
x_{R},x\right) \right).
\end{equation}
This expression is in the form suitable for extension to the forbidden
region, that is, to the right of $x_{R}$. Thus, by using (\ref{wkb9}) we
have to the right of $x_{R}$
\begin{equation}
\psi _{E}\left( x\right) \sim 2K_{E}\left[ \cos \left( \frac{1}{\hbar }%
S\left( x_{R},x_{L}\right) \right) F_{+}\left( x,x_{R}\right) +\frac{1}{2}%
\sin \left( \frac{1}{\hbar }S\left( x_{R},x_{L}\right) \right) F_{-}\left(
x,x_{R}\right) \right] ,  \label{eq89}
\end{equation}
which can be rewritten again as
\begin{equation}
\psi _{E}\left( x\right) \sim 2K_{E}\left[ \cos\left(\frac{1}{\hbar} S\left(
x_{R},x_{L}\right) \right) e^{S\left( x_{out},x_{R}\right)/\hbar
}F_{-}\left( x_{out},x\right) +\frac{1}{2}\sin\left( \frac{1}{\hbar} S\left(
x_{R},x_{L}\right) \right) e^{-S\left( x_{out},x_{R}\right)/\hbar
}F_{+}\left( x_{out},x\right) \right],
\end{equation}
which is in a form suitable for extension to the right of $x_{out}$:
\begin{eqnarray}
\psi _{E}\left( x\right) &\sim &\frac{2K_{E}}{\sqrt{2p\left( x\right)/\hbar }%
}\left[ 2 \cos \left(\frac{1}{\hbar} S\left( x_{R},x_{L}\right) \right)
e^{S\left( x_{out},x_{R}\right)/\hbar } \sin \left(\frac{1}{\hbar} S\left(
x,x_{out}\right) +\frac{\pi }{4}\right) \right.  \nonumber \\
&&\left. +\frac{1}{2} \sin\left(\frac{1}{\hbar} S\left( x_{R},x_{L}\right)
\right) e^{-S\left( x_{out},x_{R}\right)/\hbar } \cos \left(\frac{1}{\hbar}
S\left( x,x_{out}\right) +\frac{\pi }{4}\right)\right].  \label{wkb10}
\end{eqnarray}

Note that if we impose the Bohr-Sommerfeld quantization rule
\begin{equation}
S\left( x_{R},x_{L}\right) =\frac{\pi}{2}(1+2n)\hbar,  \label{wkb11}
\end{equation}
only the subdominant, exponential decreasing part survives. This would
correspond to the case when the far right region is forbidden and may be
used to define energies for false states trapped into the potential well, in
particular $n=0$ will correspond to the false ground state.

\subsection{Normalization}

All that remains now is the determination of the normalization constant $%
K_{E}$ which can be done from the normalization of the wave functions. The
eigenfunctions are subject to continuous normalization
\begin{equation}
\int dx\;\psi _{E_{1}}\left( x\right) \psi _{E_{2}}\left( x\right) =\delta
\left( E_{1}-E_{2}\right) .  \label{norm1}
\end{equation}
Since the functions themselves are regular, the singular behavior must come
from the upper limit, see for instance \cite{LanLif77}. For large enough $x,$
we have from Eq. (\ref{wkb3})
\begin{equation}
p\rightarrow p_{\infty }=\sqrt{2M\left( E+U_{\infty }\right) }.
\label{norm2}
\end{equation}

Let us write from Eq. (\ref{wkb4}),
\begin{equation}
S\left( x,x_{out}\right) =p_{\infty }\left( x-x_{out}\right)
+\int_{x_{out}}^{x}dx^{\prime }\;\left[ \sqrt{2M\left( E-U\left( x^{\prime
}\right) \right) }-\sqrt{2M\left( E+U_{\infty }\right) }\right] ,
\label{norm2a}
\end{equation}
if this integral converges, we may take the upper limit of integration to $%
\infty ,$ whereby
\begin{equation}
S\left( x,x_{out}\right) =p_{\infty }x+f\left( E\right) ,  \label{norm3}
\end{equation}
where $f(E)$ stands for the second term of (\ref{norm2a}). Then, for $x\gg
x_{out}$, we can write from (\ref{wkb10}) and (\ref{norm3})
\begin{equation}
\psi _{E}\left( x\right) \sim \frac{\sqrt{2\hbar }K_{E}}{\sqrt{p_{\infty }}}%
\left[ A\left( E\right) \sin \left( \frac{p_{\infty }x}{\hbar }\right)
+B\left( E\right) \cos \left( \frac{p_{\infty }x}{\hbar }\right) \right] ,
\label{norm4}
\end{equation}
where $A(E)$ and $B(E)$ are given by
\begin{eqnarray}
A\left( E\right)  &=&2\cos \left( \frac{1}{\hbar }S\left( x_{R},x_{L}\right)
\right) e^{S\left( x_{out},x_{R}\right) /\hbar }\cos \left( f\left( E\right)
+\frac{\pi }{4}\right)   \nonumber \\
&&-\frac{1}{2}\sin \left( \frac{1}{\hbar }S\left( x_{R},x_{L}\right) \right)
e^{-S\left( x_{out},x_{R}\right) /\hbar }\sin \left( f\left( E\right) +\frac{%
\pi }{4}\right) ,  \label{norm5a}
\end{eqnarray}
\begin{eqnarray}
B\left( E\right)  &=&2\cos \left( \frac{1}{\hbar }S\left( x_{R},x_{L}\right)
\right) e^{S\left( x_{out},x_{R}\right) /\hbar }\sin \left( f\left( E\right)
+\frac{\pi }{4}\right)   \nonumber \\
&&+\frac{1}{2}\sin \left( \frac{1}{\hbar }S\left( x_{R},x_{L}\right) \right)
e^{-S\left( x_{out},x_{R}\right) /\hbar }\cos \left( f\left( E\right) +\frac{%
\pi }{4}\right) .  \label{norm5b}
\end{eqnarray}

Substituting Eq. (\ref{norm4}) into (\ref{norm1}), the singular terms in the
normalization integral are
\[
\int dx\;\psi _{E_{1}}\left( x\right) \psi _{E_{2}}\left( x\right) \sim
\hbar^2 \pi \frac{K_{E_{1}}^{2}}{p_{ 1}}\left[ A^{2}\left( E_{1}\right)
+B^{2}\left( E_{1}\right) \right] \left[ \frac{dp_{1}}{dE_{1}}\right]
^{-1}\delta \left( E_{1}-E_{2}\right),
\]
where the delta function comes from the $x$ integration which brings $\delta
(p_{ 1}-p_{ 2})$ and where we have defined $p_{i}\equiv p_{\infty}(E_i)$ ($%
i=1,2$) and changed from momentum to energy variables according to $p_{
i}dp_{ i}=M dE_{i}$; see Eq. (\ref{norm2}). The normalization condition
reduces to $1$ the coefficient of the delta function above
\begin{equation}
\hbar ^{2}\pi \frac{K_{E_{1}}^{2}}{M}\left[ A^{2}\left( E_{1}\right)
+B^{2}\left( E_{1}\right) \right] =1.  \label{norm6}
\end{equation}
This suggests the introduction of the phase $\delta_E$ as follows,
\begin{equation}
K_{E}A\left( E\right) =\sqrt{\frac{M}{\hbar ^{2}\pi }}\cos \delta _{E},
\qquad K_{E}B\left( E\right) =\sqrt{\frac{M}{\hbar ^{2}\pi }}\sin \delta
_{E}.  \label{norm7}
\end{equation}
Thus, the eigenfunction at $x\gg x_{out}$ is Eq. (\ref{norm8}), that is
\[
\psi _{E}\left( x\right) \sim \sqrt{\frac{2M}{\hbar \pi p_{\infty }}}\sin
\left(\frac{ p_{\infty }x}{\hbar}+\delta _{E}\right).
\]

To work out the constant $K_{E}$ in greater detail we note that form Eqs. (%
\ref{norm5a}) and (\ref{norm5b}) we have
\begin{equation}
A^{2}+B^{2}=4 \cos ^{2}\left(\frac{1}{\hbar}S\left( x_{R},x_{L}\right)
\right) e^{2S\left( x_{out},x_{R}\right)/\hbar }+\frac{1}{4} \sin^2\left(%
\frac{1}{\hbar} S\left( x_{R},x_{L}\right) \right) e^{-2S\left(
x_{out},x_{R}\right)/\hbar },  \label{norm9}
\end{equation}
which is non vanishing as long as $E$ is real. However, if we allow for
complex energies, as is typical of unstable states, it may be zero provided
\begin{equation}
\cos ^{2}\left(\frac{1}{\hbar}S\left( x_{R},x_{L}\right)\right) =\frac{-1}{16%
} \sin ^{2}\left(\frac{1}{\hbar}S\left( x_{R},x_{L}\right) \right)
e^{-4S\left( x_{out},x_{R}\right)/\hbar }.  \label{norm10}
\end{equation}
The left hand side is zero whenever the energy satisfies the Bohr-Sommerfeld
condition (\ref{wkb11}).

\section{Thermal spectrum}

\label{aa}

In this Appendix we check that the quantum transport equation (\ref{open5})
admits a stationary solution with a thermal spectrum. This can be seen as a
test on the restrictions satisfied by the matrix elements (\ref{open7a})-(%
\ref{open7d}) with $\sigma^2=k_B T$.

An unnormalized thermal density matrix in the position representation reads,
\begin{equation}
\rho \left( x,x^{\prime }\right) =\int dE\;e^{-\beta E}\psi _E\left(
x\right) \psi _E\left( x^{\prime }\right),  \label{a1}
\end{equation}
where $\beta=(k_B T)^{-1}$ and its associated Wigner function is
\begin{equation}
W_\beta \left( x,p\right) =\int dE\;e^{-\beta E}W_{EE}\left( x,p\right),
\label{a2}
\end{equation}
which in the energy representation in the base $W_{E_1 E_2}$ of Eq. (\ref
{ener5}) corresponds to the coefficients $C_{E_1E_2}=e^{-\beta E_1}\delta
\left( E_1-E_2\right) $. Inserting this into the transport equation we get
\begin{equation}
\int dE\;e^{-\beta E}Q_{E_1E_2,EE}=0,  \label{a3}
\end{equation}
which after using Eqs. (\ref{open6a}), (\ref{open8a}) and (\ref{open8b}) can
be written in operator language as
\begin{eqnarray*}
0 &=&\frac 1{2M}\frac i\hbar \left( XPe^{-\beta H}-e^{-\beta H}PX-
Pe^{-\beta H}X+Xe^{-\beta H}P \right) \\
&&+\frac 1{\beta \hbar ^2}\left( X^2e^{-\beta H}+e^{-\beta H}X^2-2Xe^{-\beta
H}X\right).
\end{eqnarray*}

At the infinite temperature limit, $\beta =0,$ this is
\begin{equation}
0 =\frac 1M\frac i\hbar \left( XP-PX\right) -\frac 1{\hbar ^2}\left(
X^2H+HX^2-2XHX\right).
\end{equation}
The first term is the commutator which gives $-M^{-1}$, and the second term
can be written as $-\frac 1{\hbar ^2}\left[ X,\left[ X,H\right] \right]$,
which using $\left[ H,X\right] =( \hbar/i ) (P/M)$ is easily seen to cancel
the first term.

\section{Quantum transport equation}

\label{aaa3}

Here we write explicitly the quantum transport equation (\ref{open5}) in the
energy representation. The coefficient $Q$ in Eq. (\ref{open5}) is given by (%
\ref{open6a}), and the values of the dissipative and noise parts of this
coefficient are given, respectively, by Eqs. (\ref{open8a}) and (\ref{open8b}%
). These parts can be directly written using the matrix elements deduced in
Section \ref{ab}. When the coefficients $C_{p_{1}p_{2}}$ defined in Eq. (\ref
{ab7}) are introduced the transport equation becomes,
\begin{eqnarray}
\frac{\partial C_{p_{1}p_{2}}}{\partial t} &=&\frac{-i}{2M\hbar }%
(p_{1}^{2}-p_{2}^{2})C_{p_{1}p_{2}}+\frac{\gamma }{2}\left( \frac{\partial }{%
\partial p_{1}}+\frac{\partial }{\partial p_{2}}\right) \left[
(p_{1}+p_{2})C_{p_{1}p_{2}}\right]   \nonumber \\
&&+\frac{\gamma }{4\pi ^{2}}\frac{\partial }{\partial p_{1}}\int
dp_{1}^{\prime }dp_{2}^{\prime }\;\left( p_{2}+p_{2}^{\prime }\right)
P(p_{2}-p_{2}^{\prime })P(p_{1}-p_{1}^{\prime })C_{p_{1}^{\prime
}p_{2}^{\prime }}  \nonumber \\
&&+\frac{\gamma }{4\pi ^{2}}\frac{\partial }{\partial p_{2}}\int
dp_{1}^{\prime }dp_{2}^{\prime }\;\left( p_{1}+p_{1}^{\prime }\right)
P(p_{2}-p_{2}^{\prime })P(p_{1}-p_{1}^{\prime })C_{p_{1}^{\prime
}p_{2}^{\prime }}  \nonumber \\
&&+\frac{\gamma }{4\pi }\left( \frac{\partial \delta _{1}}{\partial p_{1}}-%
\frac{\partial \delta _{2}}{\partial p_{2}}\right) \int dp\;\left[ \left(
p_{1}+p\right) P(p_{1}-p)C_{pp_{2}}-\left( p_{2}+p\right)
P(p_{2}-p)C_{p_{1}p}\right]   \nonumber \\
&&+\gamma M\sigma ^{2}\left[ \frac{\partial ^{2}}{\partial p_{1}^{2}}+\frac{%
\partial ^{2}}{\partial p_{2}^{2}}-\left( \frac{\partial \delta _{1}}{%
\partial p_{1}}-\frac{\partial \delta _{2}}{\partial p_{2}}\right)
^{2}\right] C_{p_{1}p_{2}}  \nonumber \\
&&+\gamma M\sigma ^{2}\int dp_{1}^{\prime }dp_{2}^{\prime }\frac{2}{\pi ^{2}}%
\frac{\partial P\left( p_{1}-p_{1}^{\prime }\right) }{\partial p_{1}}\frac{%
\partial P\left( p_{2}-p_{2}^{\prime }\right) }{\partial p_{2}}%
C_{p_{1}^{\prime }p_{2}^{\prime }}  \nonumber \\
&&+\gamma M\sigma ^{2}\int \frac{dp_{2}^{\prime }}{\pi }\left( \frac{%
\partial \delta _{2}}{\partial p_{2}}+\frac{\partial \delta _{2^{\prime }}}{%
\partial p_{2}^{\prime }}-2\frac{\partial \delta _{1}}{\partial p_{1}}%
\right) \frac{\partial P(p_{2}-p_{2}^{\prime })}{\partial p_{2}}%
C_{p_{1}p_{2}^{\prime }}  \nonumber \\
&&+\gamma M\sigma ^{2}\int \frac{dp_{1}^{\prime }}{\pi }\left( \frac{%
\partial \delta _{1}}{\partial p_{1}}+\frac{\partial \delta _{1^{\prime }}}{%
\partial p_{1}^{\prime }}-2\frac{\partial \delta _{2}}{\partial p_{2}}%
\right) \frac{\partial P(p_{1}-p_{1}^{\prime })}{\partial p_{1}}%
C_{p_{1}^{\prime }p_{2}},  \label{ab8}
\end{eqnarray}
where we have used the shorthand notation $P(x)\equiv PV(1/x)$.

This equation simplifies considerably if we assume that $C_{p_1p_2}\sim
C\left( p_1\right) \delta \left( p_1-p_2\right)$, with $C\left( p\right) $
slowly varying. This is justified by noticing the effect of the second local
$\sigma^2$ term which is negative. This term has no effect on the diagonal
terms, when $p_1=p_2$, but its effect on the off-diagonal coefficients is
very important. In fact, it exponentially reduces the coefficients $C_{p_1
p_2}$ on a time scale of the decoherence time, as discussed in Section \ref
{ab}. One may argue that since tunneling is a long time process with a
typical scale of time $\tau_{\mathrm{tunn}}\sim \hbar/\varepsilon$ the local
approximation should give a reasonable approximation to the transport
equation (\ref{ab9}) whenever $\tau_D\ll \tau_{\mathrm{tunn}}$.

Now, from Eq. (\ref{ab4}) and the diagonality assumption for the $%
C_{p_{1}p_{2}}$ coefficients we have that
\[
\int dp_{1}^{\prime }dp_{2}^{\prime }\frac{1}{\pi ^{2}}\frac{\partial
P\left( p_{1}-p_{1}^{\prime }\right) }{\partial p_{1}}\frac{\partial P\left(
p_{2}-p_{2}^{\prime }\right) }{\partial p_{2}}C_{p_{1}^{\prime
}p_{2}^{\prime }}=\frac{\partial ^{2}}{\partial p_{1}\partial p_{2}}%
C_{p_{1}p_{2}}.
\]
This term together with the two local terms involving second order
derivatives of the momenta in Eq. (\ref{ab8}) become
\[
\gamma M\sigma ^{2}\left( \frac{\partial }{\partial p_{1}}+\frac{\partial }{%
\partial p_{2}}\right) ^{2}C_{p_{1}p_{2}},
\]
from where Eq. (\ref{ab9}) follows as the local approximation of the quantum
transport equation (\ref{ab8}).

\section{Tunneling rates for the closed system}

In this Appendix we review the calculation of the quantum
mechanical tunneling rate for the closed system, that is, ignoring
the interaction with the environment. The quantum tunneling rate
as given by the instanton calculation
\cite{CalLeg83b,GraWei84,MarDevCla87} is
\begin{equation}
\Gamma _{closed}^{(inst)}=\frac{a_{q}}{2\tau }e^{-\Lambda _{0}},  \label{7.7}
\end{equation}
where $\tau =\pi /\Omega _{0}$, $\Lambda _{0}=S_{B}/\hbar
=18\varepsilon
_{s}/(5\varepsilon _{0})\sim 12.\,\allowbreak 376$, and the prefactor $%
a_{q}=(120\pi \Lambda _{0})^{1/2}\sim 68.306$ . With these values,
the escape temperature defined in Eq. (\ref{tesc}) is
\begin{equation}
T_{esc}^{(inst)}= \frac{ \varepsilon _{0}/k_{B} }{3.6-
(\varepsilon _{0}/\varepsilon _{s})\ln  a_{q} }\sim 72.345\;%
\mathrm{mK}.
\end{equation}

It is interesting to check that this result agrees with the result
we obtain when the dissipation is zero. We can use our WKB result
as obtained in Sec. \ref{sec2}, see Eq. (\ref{tunn3}), to write
\begin{equation}
\Gamma _{closed}^{(WKB)}=\frac{1}{2\tau }e^{-\Lambda },  \label{7.9}
\end{equation}
where $\Lambda =(2/\hbar)S_{0}(x_{out},x_{R})$, with $S_{0}$
defined in Eqs. (\ref{wkb2})-(\ref{wkb3}), where the potential
$U(x)$ is given by Eq. (\ref{7.3}).

For a cubic potential, the relationship among the energy $E,$ the
frequency $\Omega $ and the action $S\left( x_{R},x_{L}\right) $
is best given in parametric form,
\begin{equation}
E=2\varepsilon _{s}\zeta \left( k\right),\quad \Omega =\Omega
_{0}f\left( k\right),\quad S\left( x_{R},x_{L}\right)
=\frac{\varepsilon _{s}}{\Omega _{0}}F\left( k\right), \label{D4}
\end{equation}
with $0<k<1$, and
\begin{eqnarray}
&&\zeta \left( k\right) =\frac{1}{8}\left\{ 2+3\frac{\left( 1+k^{2}\right) }{%
\left[Q\left( k\right)\right]^{1/2} }
-\frac{\left( 1+k^{2}\right) ^{3}}{\left[
Q\left( k\right) \right] ^{3/2}}\right\},  \label{D7}\\
&&f\left( k\right) =\left\{ \frac{2}{\pi }\left[ 4Q\left( k\right)
\right] ^{1/4}K\left[ k^{2}\right] \right\} ^{-1},\nonumber\\
&&F\left( k\right) =\frac{27}{8}\left[ \frac{4}{Q\left( k\right)
}\right] ^{5/4}\left\{ a\left( k\right) E\left[ k^{2}\right]
-\left( 1-k^{2}\right) b\left( k\right) K\left[ k^{2}\right]
\right\},  \label{D9}
\end{eqnarray}
where $E\left[ k^{2}\right] $ and $K\left[ k^{2}\right] $ are the
complete elliptic integrals, and we have introduced the functions
$ Q\left( k\right) =(1/4)\left( 1+14k^{2}+k^{4}\right),$ $a\left(
k\right) =(16/15)\left( 2-k^{2}\right) ^{2}-(1/5)\left(
1-k^{2}\right) \left( 21-5k^{2}\right)$ and $b\left( k\right)
=(8/15)\left( 2-k^{2}\right) -\left( 1-k^{2}\right).$

The Bohr-Sommerfeld condition Eq. (\ref{wkb11}) for the ground
state ($n=0$), corresponds to the parameter $k_{GS}$ such that $
F\left( k_{GS}\right) =\pi \varepsilon _{0}/\varepsilon _{s}$
which implies  that $k_{GS}\sim 0.1152$. This corresponds to
$\zeta \left( k_{GS}\right) \sim 0.1423$ and $f\left(
k_{GS}\right) \sim 0.9550,$ while the harmonic approximation for
the potential yields $0.1454$ and $1$, respectively.

To compute the barrier penetrability,
$\Lambda=(2/\hbar)S_0(x_{out},x_R)$, we observe that
$S\left( x_{out},x_{R}\right) $ at energy $E$ is equal to
$S\left( x_{R},x_{L}\right) $ at energy $E_{ref}=\varepsilon
_{s}-E.$ The exchange of $E$ by
$E_{ref}$ is equivalent to the exchange of $k$ by $k_{ref},$ where $%
\zeta \left( k_{ref}\right) = 1/2 -\zeta \left( k\right) .$
For 
$k_{GS}$ we obtain $k_{ref}\sim 0.2433$ and $F\left(
k_{ref}\right) \sim 2.4073$. Therefore
\begin{equation}
\Lambda =\frac{\varepsilon _{s}}{\varepsilon _{0}}F\left(
k_{ref}\right) \sim 8.459. \label{D10}
\end{equation}
This is to be compared against the instanton exponent $\Lambda
_{0}-\ln  a_{q} \sim 8.152.$ In terms of the escape temperature,
the WKB approximation yields
\begin{equation}
T_{esc}^{(WKB)}=\frac{ \varepsilon _{0}/k_{B} }{F\left(
k_{ref}\right) -(\varepsilon _{0}/\varepsilon _{s})\ln \left(
\Omega _{GS}/\Omega _{0}\right) }\sim 70. 869\;\mathrm{mK},
\end{equation}
which is in good agreement with the instanton result. This, of
course, should not be surprising since for a closed system our
method reduces to the standard WKB calculation. The purpose of
this exercise is just to check the consistency of our calculation
and to illustrate how the instanton and WKB methods compare. That
the difference between $\exp (\Lambda _{0})$ and $\exp (\Lambda )$
is accounted for by the prefactor $a_{q}$ of Eq. (\ref{7.7}) can
be seen analytically by a perturbative calculation.


\begin{thebibliography}{99}

\bibitem{CalLeg83b}  A.O. Caldeira and A.J. Leggett, Ann. Phys.
\textbf{149}, 374 (1983).

\bibitem{DevMarCla85}  M. H. Devoret, J. M. Martinis and J. Clarke, Phys.
Rev. Lett. \textbf{55}, 1908 (1985).

\bibitem{MarDevCla87}  J. M. Martinis, M. H. Devoret, and J. Clarke, Phys.
Rev. B \textbf{35}, 4682 (1987).

\bibitem{CleMarCla88}  A. N. Cleland, J. M. Martinis and J. Clarke, Phys.
Rev. B \textbf{37}, 5950 (1988).

\bibitem{WalEtAl03}  A. Wallraff, T. Duty, A. Lukashenko, and A. V. Ustinov,
Phys. Rev. Lett, \textbf{90}, 037003 (2003).

\bibitem{CalLeg81}  A.O. Caldeira and A.J. Leggett, Phys. Rev. Lett. \textbf{%
46}, 211 (1981).

\bibitem{LegEtAl87}  A.J. Leggett, S. Chakravarty, A. T. Dorsey, M. P. A.
Fisher, A. Garg and W. Zwerger, Rev. Mod. Phys. \textbf{59}, 1 (1987).

\bibitem{Han86}  P. H\"{a}nggi, Ann. N. Y. Acad. Sci. \textbf{480}, 51
(1986).

\bibitem{Han87}  P. H\"{a}nggi, Z. Phys. B \textbf{68}, 181 (1987).

\bibitem{WeiEtAl87}  U. Weiss, H. Grabert, P. H\"{a}nggi and P. S.
Riseborough, Phys. Rev. B \textbf{35}, 9535 (1987).

\bibitem{GraOlsWei87}  H. P. Grabert, P. Olschowski and U. Weiss, Phys. Rev.
B \textbf{36}, 1931 (1987).

\bibitem{GraWeiHan84}  H. Grabert, U. Weiss and P. H\"{a}nggi,
Phys. Rev. Lett. \textbf{52}, 2193 (1984).

\bibitem{GriEtAl89}  U. Griff, H. Grabert, P. H\"{a}nggi, and P. S.
Riseborough,
Phys. Rev. B \textbf{40}, 7295 (1989).

\bibitem{MarGra88}  J. M. Martinis and H. Grabert, Phys. Rev. B \textbf{38},
2371 (1988).

\bibitem{GraWei84}  H. Grabert and U. Weiss, Phys. Rev. Lett. \textbf{53},
1787 (1984).

\bibitem{Wei93}  U. Weiss, \textit{Quantum Dissipative Systems} (World
Scientific, Singapore, 1993).

\bibitem{HanTalBor90}  P. H\"{a}nggi, P. Talkner, and M. Borkovec, Rev. Mod.
Phys. \textbf{62}, 251 (1990).

\bibitem{Lan67}  J. Langer, Ann. Phys. \textbf{41}, 108 (1967).

\bibitem{VolKobOku75}  M. B. Voloshin, I. Yu Kobzarev and L.B. Okun, Sov. J.
Nucl. Phys. \textbf{20}, 644 (1975).

\bibitem{Col77}  S. Coleman, Phys. Rev. D \textbf{15}, 2929 (1977).

\bibitem{CalCol77}  C. Callan and S. Coleman, Phys. Rev. D \textbf{16}, 1762
(1977).

\bibitem{ColGlaMar78}  S. Coleman, V. Glaser and A. Martin, Comm. Math. Phys
\textbf{58}, 211 (1978).

\bibitem{ColDeL80}  S. Coleman and F. De Luccia, Phys. Rev. D \textbf{21},
3305 (1980).

\bibitem{Col85}  S. Coleman, \textit{Aspects of Symmetry} (Cambridge
University Press, Cambridge, England, 1985.

\bibitem{Kra40}  H. Kramers, Physica VII, 284 (1940).

\bibitem{ShnSchHer97}
A.Shnirman, G. Sch\"on, and Z. Hermon, Phys. Rev. Lett. \textbf{79}, 2371 (1997).

\bibitem{MooEtAl99} J.  E.  Mooij,  T.  P.  Orlando,  L.  Levitov,  L.  Tian, 
C. H.  van  der  Wal,  and  S.  Lloyd,  Science, \textbf{285}, 1036 (1999). 

\bibitem{CleGel04}
A. N. Cleland and M. R. Geller,  Phys. Rev. Lett. \textbf{93}, 070501 (2004).

\bibitem{Mon02}
C. Monroe, Nature \textbf{416}, 238 (2002).

\bibitem{Kib80}  T. W. B. Kibble, Phys. Rep. \textbf{67}, 183 (1980).

\bibitem{RivLomMaz02}  R. J. Rivers, F. C. Lombardo and F.D. Mazzitelli
Phys. Lett. B \textbf{539}, 1 (2002).

\bibitem{Lan69}  J. Langer, Ann. Phys. (N. Y.) \textbf{54}, 258 (1969).

\bibitem{CalVer99}  E.~Calzetta and E.~Verdaguer,
Phys.\ Rev.\ D \textbf{59}, 083513 (1999).

\bibitem{CalHu94}  E.~Calzetta and B.~L.~Hu, Phys.\ Rev.\ D \textbf{49},
6636 (1994).

\bibitem{CamVer96}  A.~Campos and E.~Verdaguer,
Phys.\ Rev.\ D \textbf{53}, 1927 (1996).

\bibitem{CalCamVer97}  E.~Calzetta, A.~Campos and E.~Verdaguer,
Phys.\ Rev.\ D \textbf{56}, 2163 (1997).

\bibitem{MarVer99a}  R.~Martin and E.~Verdaguer,
Phys.\ Rev.\ D \textbf{60}, 084008 (1999).

\bibitem{MarVer99b}  R.~Martin and E.~Verdaguer,
Phys.\ Lett.\ B \textbf{465}, 113 (1999).

\bibitem{HuVer03}  B.~L.~Hu and E.~Verdaguer, Class. Quantum Grav. \textbf{20%
}, R1 (2003).

\bibitem{HuVer04}  B.~L.~Hu and E.~Verdaguer, Living Rev. Relativity \textbf{%
7}, 3 (2004).

\bibitem{CalRouVer01a}  E.~Calzetta, A.~Roura and E.~Verdaguer,
Phys.\ Rev.\ D \textbf{64}, 105008 (2001).

\bibitem{CalRouVer02}  E.~Calzetta, A.~Roura and E.~Verdaguer,
Phys.\ Rev.\ Lett.\ \textbf{88}, 010403 (2002).

\bibitem{ArtEtAl03}  D.~Arteaga, E.~Calzetta, A.~Roura and E.~Verdaguer,
Int. J. Theor. Phys. \textbf{42}, 1257 (2003).

\bibitem{Ris89}  H. Risken, \textit{The Fokker-Planck Equation}
(Springer-Verlag, Berlin, 1989).

\bibitem{VogRis88}  K. Vogel and H. Risken,
Phys. Rev. A \textbf{38 }, 2409 (1988).

\bibitem{RisVog88}  H. Risken and K. Vogel, Quantum treatment of dispersive
optical bistability, in \textit{Far from equilibrium phase transitions},
Springer, Berlin (1988).

\bibitem{GarZue04}  J. L. Garc\'{\i }a-Palacios and D. Zueco, e-print
cond-mat/0407454.

\bibitem{HuPazZha92}  B.L. Hu, J.P. Paz, and Y. Zhang, Phys. Rev. D
\textbf{45}, 2843 (1992).

\bibitem{Zur91}  W.H. Zurek, Physics Today \textbf{44}, 36 (1991).

\bibitem{PazHabZur93}  J.P. Paz, S. Habib and W.H. Zurek, Phys. Rev. D
\textbf{47}, 488 (1993).

\bibitem{ZurPaz94}  W.H. Zurek, and J.P. Paz, Phys. Rev. Lett. \textbf{72},
2508 (1994); ibid. \textbf{75}, 351 (1995).

\bibitem{PazZur99}  J.P. Paz, and W.H. Zurek, Phys. Rev. Lett. \textbf{82},
5181 (1999).

\bibitem{PazZur01}  J.P. Paz, and W.H. Zurek, in \textit{Coherent Matter
Waves}, Lectures from the 72nd Les Houches Summer School, 1991, edited by R.
Kaiser, C. Westbrook, and F. David (Springer-Verlag, Berlin, 2001), pp.
533-614.

\bibitem{MonPaz00}  D. Monteoliva and J. P. Paz, Phys. Rev. Lett. \textbf{85}%
, 3373 (2000).

\bibitem{MonPaz01}  D. Monteoliva and J. P. Paz,
Phys. Rev. E \textbf{64}, 056238 (2001).

\bibitem{Tin96}  M. Tinkham, \textit{Introduction to superconductivity},
McGraw-Hill, New York, 1996.

\bibitem{LanLif77}  L. D. Landau and E. M. Lifshitz, \textit{Quantum
mechanics} (third edition), Butterworth-Heinemann, Oxford, 1977.

\bibitem{GalPas90}  A. Galindo and P. Pascual, \textit{Quantum mechanics,
vol. 2}, Springer-Verlag, Heidelberg, 1990.

\bibitem{Mig77}  A. Migdal, \textit{Qualitative methods in quantum
Mechanics, }Addison-Wesley, New York, 1977.

\bibitem{Wig32}  E. P. Wigner, Phys. Rev. \textbf{40}, 749 (1932).

\bibitem{HilEtAl84}  M. Hillery, R. F. O'Connell, M. O. Scully and E. P.
Wigner, Phys. Rep. \textbf{106}, 121 (1984).

\bibitem{Shi65}  J. Shirley, 
Phys. Rev. A \textbf{138} 979 (1965).

\bibitem{MilWya83}  K.F. Milfeld and R. Wyatt, 
Phys. Rev. A 27, 72 (1983).

\bibitem{UteDitHan94}  R. Utermann, T. Dittrich, and P. H\"{a}nggi,
Phys. Rev. E \textbf{49}, 273 (1994).

\bibitem{BluEtAl91}  R. Blumel, A. Buchleitner, R. Graham, L. Sirko, U.
Smilansky, and H. Walther, 
Phys. Rev. A \textbf{44}, 4521 (1991).

\bibitem{CalRouVer03}  E.~Calzetta, A.~Roura and E.~Verdaguer,
Physica A \textbf{319}, 188 (2003).

\bibitem{CalLeg83a}  A.O. Caldeira and A.J. Leggett, Physica A \textbf{121},
587 (1983).

\bibitem{UnrZur89}  W. G. Unruh and W. H. Zurek,
Phys. Rev. D \textbf{40}, 1071 (1989).

\bibitem{HuPazZha93}  B.L. Hu, J.P. Paz, and Y. Zhang, Phys. Rev. D \textbf{%
47}, 1576 (1993).

\bibitem{HalYu96}  J. J. Halliwell and T. Yu, Phys. Rev. D \textbf{53}, 2012
(1996).

\bibitem{AntLomMon01}  N. D. Antunes, F. C. Lombardo and D. Monteoliva Phys.
Rev. E \textbf{64}, 066118 (2001).

\bibitem{FulDun74} T. A. Fulton and L. N. Dunkleberger,
Phys. Rev. B \textbf{9}, 4760 (1974).

\bibitem{SchEtAl85}
D. B. Schwartz, B. Sen, C. N. Archie and J. E. Lukens, 
Phys. Rev. Lett. \textbf{55}, 1547 (1985).

\bibitem{SilLieGra88}
P. Silvestrini, O. Liengme and K. E. Gray, Phys. Rev. B \textbf{37},
1525 (1988).

\bibitem{Sil92}
P. Silvestrini, Phys. Rev. B \textbf{46}, 5470 (1992).

\bibitem{LiEtAl02}
S-X. Li, Y. Yu, Y. Zhang, W. Qiu,  S. Han and Z. Wang,
Phys. Rev. Lett. \textbf{89}, 98301 (2002).

\bibitem{MonMygRiv02}
R. Monaco, J. Mygind and R. J. Rivers, Phys. Rev. Lett \textbf{89},
80603 (2002).

\bibitem{MonMygRiv03}
R. Monaco, J. Mygind and R. J. Rivers, Phys. Rev. B \textbf{67},
104506 (2003).

\bibitem{FisWalUst03}
M. V. Fistful, A. Wallraff and A. V. Ustinov, Phys. Rev. B \textbf{68},
60504 (2003).

\bibitem{GraOlsWei86}  H. Grabert, P. Olschowski and U. Weiss, Phys. Rev.
Lett \textbf{57}, 265 (1986).

\end{thebibliography}
\end{document}